\documentclass[10pt,twocolumn]{article}
\usepackage[top=1.8cm, bottom=1.8cm, left=1.8cm, right=1.8cm]{geometry}
\usepackage{helvet}  %
\usepackage{courier}  %
\usepackage[hyphens]{url}  %
\usepackage[keeplastbox]{flushend}  %
\usepackage{graphicx}  %
\newif\ifcomment
 \commentfalse

\usepackage{xcolor}
\usepackage{booktabs}
\usepackage{graphicx}
\usepackage{paralist}
\usepackage[small,bf]{caption}
\usepackage{subfigure}
\usepackage{times}
\usepackage[hang,flushmargin]{footmisc}

\usepackage[compact]{titlesec}
\titlespacing*{\section}{0pt}{*4}{4pt} 
\titlespacing{\subsection}{0pt}{*3}{3pt}

\usepackage[bookmarks=true,%
            bookmarksnumbered=true,%
            colorlinks=true,%
            linkcolor=linkcol,%
            citecolor=citecol,%
            urlcolor=urlcol,%
            hypertexnames=true,%
            pdfpagelabels]%
      {hyperref}

\definecolor{linkcol}{rgb}{0,0,0.5}
\definecolor{citecol}{rgb}{0,0.5,0.3}
\definecolor{urlcol}{rgb}{0.3,0,0}

\usepackage{xspace}

\usepackage{ifthen}
\newif\ifshort
\shortfalse %
\ifshort
  \newcommand{\isShort}{true}
\else
  \newcommand{\isShort}{false}
\fi
\newcommand{\shortVer}[1]{\ifthenelse{\equal{\isShort}{true}}{{#1}}{}}
\newcommand{\longVer}[1]{\ifthenelse{\equal{\isShort}{false}}{{#1}}{}}

\newif\ifcomment
\commenttrue
\ifcomment
\newcommand{\sz}[1]{{\bf \textcolor{blue}{#1}}}
\newcommand{\jbnote}[1]{{\bf \textcolor{magenta}{JB: #1}}}
\newcommand{\jf}[1]{{\bf \textcolor{red}{JF: #1}}}

\else
\newcommand{\sz}[1]{}
\newcommand{\jbnote}[1]{}
\newcommand{\jf}[1]{}
\fi

\newcommand{\descr}[1]{\smallskip\noindent\textbf{#1}}

\renewenvironment{thebibliography}[1]{
  \begin{oldthebibliography}{#1}
    \setlength{\itemsep}{0.1em}
    \setlength{\parskip}{0.1em}
}
{
  \end{oldthebibliography}
}

\renewcommand{\footnoterule}{%
  \kern -3pt
  \hrule width 1in 
  \kern 2pt
}

\newcommand{\dspol}{{/pol/}\xspace}

\newcommand{\td}{The\textunderscore Donald\xspace}

\usepackage{url}
\makeatletter
\def\url@leostyle{%
  \@ifundefined{selectfont}{\def\UrlFont{}}%
  {\def\UrlFont{}}%
}
\makeatother
\definecolor{darkred}{RGB}{153,0,0}
\definecolor{darkblue}{RGB}{0,0,119}
\hypersetup{colorlinks=true, linkcolor = darkred,   citecolor = darkred, urlcolor = black}

\usepackage{etoolbox}
\makeatletter
\patchcmd\@combinedblfloats{\box\@outputbox}{\unvbox\@outputbox}{}{%
   \errmessage{\noexpand\@combinedblfloats could not be patched}%
}%
 \makeatother
\AtBeginShipout{%
  \ifnum\value{page}>1 %
    \typeout{* Additional boxing of page `\thepage'}%
    \setbox\AtBeginShipoutBox=\hbox{\copy\AtBeginShipoutBox}%
  \fi
}

\begin{document}
\title{\bf A Quantitative Approach to Understanding Online Antisemitism\thanks{To appear at the 14th International AAAI Conference on Web and Social Media (ICWSM 2020) -- please cite accordingly. Work done while first author was with Cyprus University of Technology.}}
\author{Savvas Zannettou$^\dagger$, Joel Finkelstein$^{\star+}$, Barry Bradlyn$^{\diamond}$, Jeremy Blackburn${^\ddagger}$\\[0.3ex]
\normalsize $^\dagger$Max Planck Institute for Informatics, $^{\star}$Network Contagion Research Institute, $^{^{+}}$Princeton University\\
 \normalsize   ${^\diamond}$University of Illinois at Urbana-Champaign, ${^\ddagger}$Binghamton University\\
\normalsize szannett@mpi-inf.mpg.de, joel@ncri.io, bbradlyn@illinois.edu, jblackbu@binghamton.edu
}
\date{}

\maketitle

\begin{abstract}
A new wave of growing antisemitism, driven by fringe Web communities, is an increasingly worrying presence in the socio-political realm.
The ubiquitous and global nature of the Web has provided tools used by these groups to spread their ideology to the rest of the Internet.
Although the study of antisemitism and hate is not new, the scale and rate of change of online data has impacted the efficacy of traditional approaches to measure and understand these troubling trends.

In this paper, we present a large-scale, quantitative study of online antisemitism.
We collect hundreds of million posts and images from alt-right Web communities like 4chan's Politically Incorrect board (\dspol) and Gab.
Using scientifically grounded methods, we quantify the escalation and spread of antisemitic memes and rhetoric across the Web.
We find the frequency of antisemitic content greatly increases (in some cases more than doubling) after major political events such as the 2016 US Presidential Election and the ``Unite the Right'' rally in Charlottesville.
We extract semantic embeddings from our corpus of posts and demonstrate how automated techniques can discover and categorize the use of antisemitic terminology.
We additionally examine the prevalence and spread of the antisemitic ``Happy Merchant'' meme, and in particular how these fringe communities influence its propagation to more mainstream communities like Twitter and Reddit.
Taken together, our results provide a data-driven, quantitative framework for understanding online antisemitism.
Our methods serve as a framework to augment current qualitative efforts by anti-hate groups, providing new insights into the growth and spread of hate online.

\end{abstract}

\section{Introduction}\label{sec:intro}

\noindent With the ubiquitous adoption of social media, online communities have played an increasingly important role in the real-world.
The news media is filled with reports of the sudden rise in nationalistic politics coupled with racist ideology~\cite{sunstein2018republic} generally attributed to the loosely defined group known as the alt-right~\cite{splcaltright}, a movement that can be characterized by the relative youth of its adherents and relatively transparent racist ideology~\cite{adl_altright_racist}.
The alt-right differs from older groups primarily in its use of online communities to congregate, organize, and disseminate information in weaponized form~\cite{marwick2017media}, often using humor and taking advantage of the scale and speed of communication the Web makes possible~\cite{flores2018mobilizing,hine2017kek,zannettou2017web,zannettou2018gab,zannettou2018origins,morstatter2018alt,zannettou2018understanding}.
Recently, these fringe groups have begun to weaponize digital information on social media~\cite{zannettou2017web}, in particular the use of weaponized humor in the form of memes~\cite{zannettou2018origins}.

While the online activities of the alt-right are cause for concern, this behavior is not limited to the Web:
there has been a recent spike in hate crimes in the United States~\cite{cshe2018hate}, a general proliferation of fascist and white power groups~\cite{splc2017year}, a substantial increase in white nationalist propaganda on college campuses~\cite{adl2018white}.
This worrying trend of real-world action mirroring online rhetoric indicates the need for a better understanding of online hate and its relationship to real-world events.

Antisemitism in particular is seemingly a core tenet of alt-right ideology, and has been shown to be strongly related to authoritarian tendencies not just in the US, but in many countries~\cite{dunbar2003individual,frindte2005old}.
Historical accounts concur with these findings: antisemitic attitudes tend to be used by authoritarian ideologies in general~\cite{adorno1950authoritarian,arendt1973origins}.
Due to its pervasiveness, historical role in the rise of ethnic and political authoritarianism, and the recent resurgence of hate crimes, understanding online antisemitism is of dire import.
Although there are numerous anecdotal accounts, we lack a clear, large-scale, quantitative measurement and understanding of the scope of online semitism, and how it spreads between Web communities.

The study of antisemitism and hate, as well as methods to combat them are not new.
Organizations like the Anti-Defamation League (ADL) and the Southern Poverty Law Center (SPLC) have spent decades attempting to address this societal problem.
However, these organizations have traditionally taken a qualitative approach, using surveys and a relatively small number of subject matter experts to manually examine content deemed hateful.
While these techniques have produced many valuable insights, qualitative approaches are extremely limited considering the ubiquity and scale of the Web.

In this paper, we take a different approach.
We present an open, scientifically rigorous framework for quantitative analysis of online antisemitism.
Our methodology is transparent and generalizable, and our data will be made available upon request.
Using this approach, we characterize the rise of online antisemitism across several axes.
More specifically we answer the following research questions:
\begin{compactenum}
  
  \item \textbf{RQ1:} Has there been a rise in online antisemitism, and if so, what is the trend?
  \item \textbf{RQ2:} How is online antisemitism expressed, and how can we automatically discover and categorize newly emerging antisemitic language?
  \item \textbf{RQ3:} To what degree are fringe communities influencing the rest of the Web in terms of spreading antisemitic propaganda?

\end{compactenum}

\begin{figure}[t]
\centering
\subfigure[]{\includegraphics[width=0.42\columnwidth]{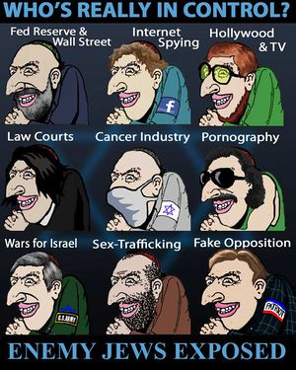}\label{fig:merchant_example_1}}
\subfigure[]{\includegraphics[width=0.45\columnwidth]{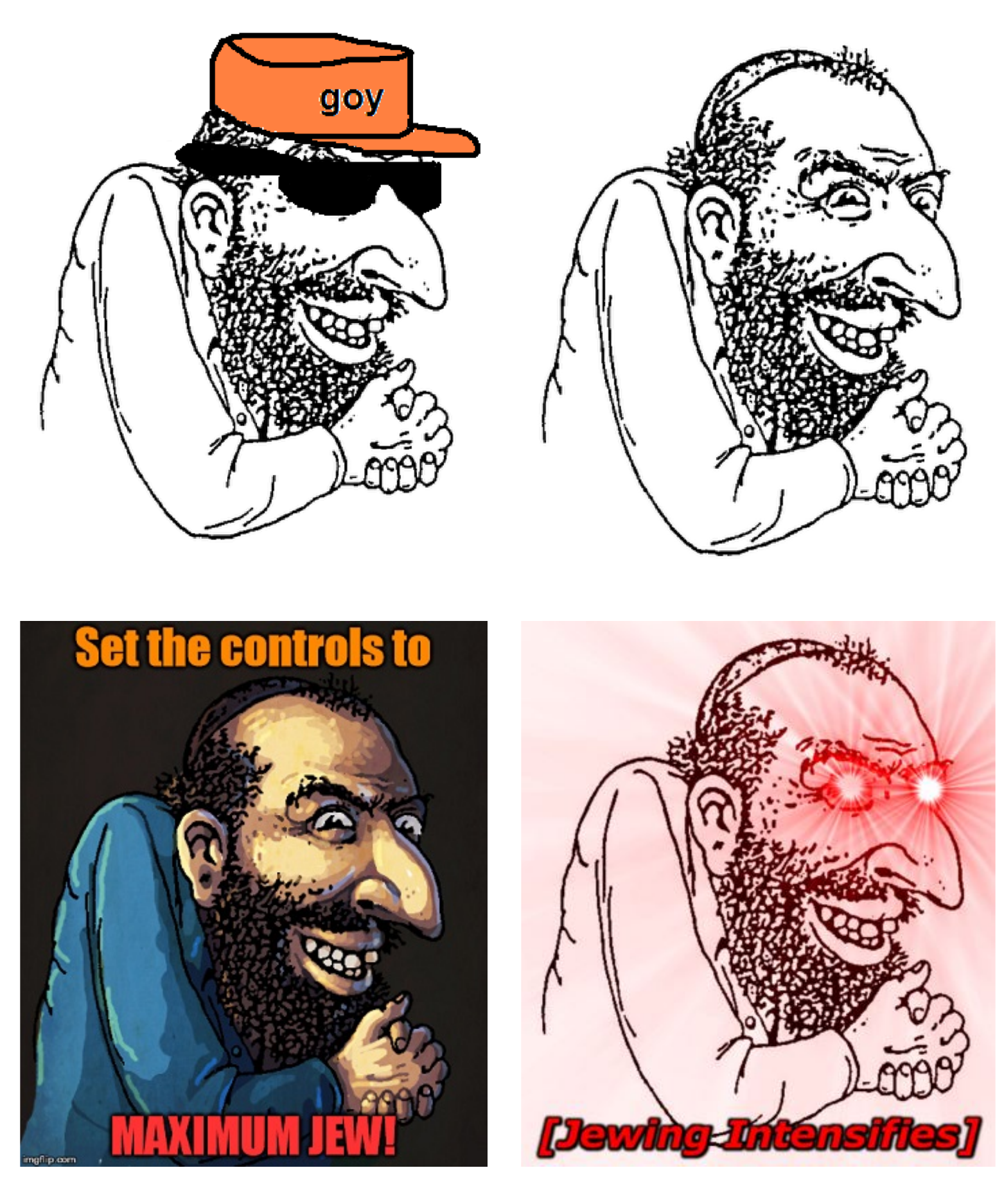}\label{fig:merchant_example_2}}
\caption{Examples of the antisemitic Happy Merchant Meme.}
\label{fig:merchant-example}
\end{figure}

We shed light to these questions by analyzing a dataset of over 100M posts from two fringe Web communities: 4chan's Politically Incorrect board (\dspol) and Gab.
We use word2vec~\cite{mikolov2013efficient} to train ``continuous bag-of-words models'' using the posts on these Web communities, in order to understand and discover new antisemitic terms.
Our analysis reveals thematic communities of derogatory slang words, nationalistic slurs, and religious hatred toward Jews.
Also, we analyze almost 7M images using an image processing pipeline proposed by~\cite{zannettou2018origins} to quantify the prevalence and diversity of the notoriously antisemitic Happy Merchant meme~\cite{happy_merchant_meme} (see Fig.~\ref{fig:merchant-example}).
We find that the Happy Merchant enjoys substantial popularity in both communities, and its usage overlaps with other general purpose (i.e.~not intrinsically antisemitic) memes.
Finally, we use Hawkes Processes~\cite{hawkes1971spectra} to model the relative influence of several fringe and mainstream communities with respect to dissemination of the Happy Merchant meme.

\descr{Disclaimer.} Note that content posted on both Web communities can be characterized as offensive and racist. In the rest of the paper, we present our analysis without censoring any offensive language, hence we inform the reader that the rest of the paper contains language and images that are likely to be upsetting. 

\section{Related Work}
In this section, we present previous related work that focus on understanding hate speech on various Web communities, detecting hate speech, and understanding antisemitism on the Web.

\descr{Hate Speech on Web Communities.} Several studies focus on understanding the degree of hate speech that exists in various Web communities.
Specifically, Hine et al.~\cite{hine2017kek} focus on 4chan's Politically Incorrect board (\dspol) by analyzing 8M posts during the course of two and a half months.
Using the Hatebase database they find that 12\% of the posts are hateful, hence highlighting \dspol's high degree of hate speech.
Similarly, Zannettou et al.~\cite{zannettou2018gab} undertake a similar analysis on Gab finding that Gab exhibits two times less the hate speech of \dspol, whereas when compared to Twitter it has two times more hateful posts.
Silva et al.~\cite{silva2016analyzing} use the Hatebase database to study hate speech on two Web communities, namely Twitter and Whisper.
Their quantitative analysis sheds light on the targets (recipients) of hate speech on the two Web communities. 
Similarly, Mondal et al.~\cite{mondal2017ams} use the same Web communities to understand the prevalence of hate speech, the effects of anonymity, as well as identify the forms of hate speech in each community.

\descr{Hate Speech Detection.}A substantial body of prior work focus on the detection of hate speech on Web communities.
Specifically, Warner and Hirschberg~\cite{warner2012detecting} use decision lists in conjunction with an SVM classifier to detect hateful content.
They evaluate the proposed approach on a classification pilot that aim to distinguish antisemitic content, highlighting that their approach has acceptable accuracy (94\%), whereas precision and recall are mediocre (68\% and 60\%, resp.)
Kwok and Wang~\cite{kwok2013locate} use a Naive Bayes classifier on tweets to classify them as either racist against blacks or non-racist.
Their classifier achieves an accuracy of 76\%, hence highlighting the challenges in discerning racist content using machine learning.
Djuric et al.~\cite{djuric2015hate} leverage a continuous bag of words (CBOW) model within doc2vec embeddings to generate low-dimensional text representations from comments posted on the Yahoo finance website. 
These representations are then fed to a binary classifier that classifies comments as hateful or not; they find that 
the proposed model outperforms BOW baselines models.

Gitari et al.~\cite{gitari2015lexicon} use subjectivity and sentiment metrics to build a hate lexicon that is subsequently used in a classifier that determines whether content is hateful.
Waseem and Hovy~\cite{waseem2016hateful} annotate 16K tweets as racist, sexist or neither.
They also assess which features of tweets contribute more on the detection task, finding that character n-grams along with a gender feature provide the best performance.
Del Vigna et al.~\cite{vigna2017hate} propose the use of Support Vector Machines (SVMs) and Recurrent Neural Networks (RNN) for the detection of hateful Italian comments on Facebook, while Ross et al.~\cite{ross2017measuring} provide a German hate speech corpus for the refugee crisis.

Serra et al.~\cite{serra2017class} use the error signal of class-based language models as a feature to a neural classifier, hence allowing to capture online behavior that uses new or misspelled words.
This approach help outperform other baselines on hate speech detection by 4\%  11\%.
Founta et al.~\cite{founta2018unified} propose the use of a unified deep learning model for the classification of tweets into different forms of hate speech like hate, sexism, bullying, and sarcasm.
The proposed model is able to perform inference on the aforementioned facets of abusive content without fine tuning, while at the same time it outperforms state-of-the-art models.

Saleem et al.~\cite{saleem2017awo} approach the problem through the lens of multiple Web communities by proposing a community-driven model for hate speech detection.
Their evaluation on Reddit, Voat, and Web forums data highlight that their model can be trained on one community and applied on another, while outperforming keyword-based approaches.
Davidson et al.~\cite{davidson2017automated} leverage the Hatebase database and crowdsourcing to annotate tweets that may contain hateful or offensive language. 
Using this dataset, they built a detection model using Logistic Regression. 
Their analysis highlights that racist and homophobic tweets are likely to be classified as hate speech, while sexist tweets are usually classified as offensive. 

Burnap and Williams~\cite{burnap2016us} propose a set of classification tools that aim to assess hateful content with respect to race, sexuality, and disability, while at the same time proposing a blended model that classifies hateful content that may contain multiple classes (e.g., race and sexuality).
Badjatiya et al.~\cite{badjatiya2017deep} compare a wide variety of machine and deep learning models for the task of detecting hate speech.
They conclude that the use of deep learning models provide a substantial performance boost when compared with character and words n-grams.

Gao et al.\cite{gao2017recognizing} propose the use of a semi-supervised approach for the detection of implicit and explicit hate speech, which mitigate costs of the annotation process and possible biases.
Also, their analysis on tweets posted around the US elections highlights the prevalence of hate on posts about the elections and the partisan nature of these posts.
In their subsequent work,  Gao and Huang~\cite{gao2017detecting} aim to tackle the hate speech detection by introducing context information on the classification process. 
Their experimental setup on news articles' comments highlights that the introduction of context information on Logistic Regression and neural networks provides a performance boost between 3\% and 7\% in terms of F1 score.

Elsherief et al.~\cite{elsherief2018peer} perform a personality analysis on instigators and recipients of hate speech on Twitter.
They conclude that both groups comprises eccentric individuals, and that instigators mainly target popular users with (possibly) a goal to get more visibility within the platform.
In their subsequent work, Elsherief et al.~\cite{Elsherief2018hate} perform a linguistic-driven analysis of hate speech on social media. 
Specifically, they differentiate hate speech in targeted hate (e.g., towards a specific individual) and generalized (e.g., towards a specific race) and find that targeted hate is angrier and more informal while generalized hate is mainly about religion.

Finally, Olteanu et al.~\cite{olteanu2017the} propose the use of user-centered metrics (e.g., users' overall perception of classification quality) for the evaluation of hate speech detection systems.

\descr{Case Studies.} Magu et al.~\cite{magu2017detecting} undertake a case study on Operation Google, a movement that aimed to use benign words in hateful contexts to trick Google's automated systems.
Specifically, they build a model that is able to detect posts that use benign words in hateful contexts and undertake an analysis on the set of Twitter users that were involved in Operation Google.
Smedt et al.~\cite{smedt2018automatic} focus on Jihadist hate speech by proposing a hate detection model using Natural Language Processing and Machine Learning techniques.
Furthermore, they undertake a quantitative and qualitative analysis on a corpus of 45K tweets and examine the users involved in Jihadist hate speech.

\descr{Antisemitism.} Leets~\cite{leets2002experiencing} surveys 120 Jews or homosexual students to assess their perceived consequences of hate speech, to understand the motive behind hate messages, and if the recipients will respond or seek support after the hate attack.
The main findings is that motives are usually enduring, that recipients respond in a passively manner while they often seek support after hate attacks.
Shainkman et al.~\cite{shainkman2016different} use the outcomes of two surveys from EU and ADL to assess how the level of antisemitism relates to the perception of antisemitism by the Jewish community in eight different EU countries.
Alietti et al.~\cite{alietti2013religious} undertake phone surveys of 1.5K Italians on islamophobic and antisemitic attitudes finding that there is an overlap of ideology for both types of hate speech.
Also, they investigate the use of three indicators (anomie, ethnocentrism, and authoritarianism) as predictors for Islamophobia and antisemitism.
Ben-Moshe et al.~\cite{benmoshe2014antisemitism} uses focus groups to explore the impact of antisemitic behavior to Jewish children. 
They conclude that there is a need for more education in matters related to racism, discrimination, and antisemitism.
Bilewicz et al.~\cite{bilewicz2013harmful} make two studies on antisemitism in Poland finding that Jewish conspiracy is the most popular and older antisemitic belief.
Furthermore, they report the personality and identity traits that are more related to antisemitic behavior.

\descr{Remarks.}  In contrast with the aforementioned work, we focus on studying the dissemination of antisemitic content on the Web by undertaking a large-scale quantitative analysis.
Our study focuses on two fringe Web communities; \dspol and Gab, where we study the dissemination of racial slurs and antisemitic memes.

\section{Datasets}\label{sec:datasets}

To study the extent of antisemitism on the Web, we collect two large-scale datasets from \dspol and Gab (see Table~\ref{tbl:datasets}).

\descr{\dspol.}
4chan is an anonymous image board that is usually exploited by troll users.
A user can create a new thread by creating a post that contains an image. 
Other users can reply below with or without images and possibly add references to previous posts.
The platform is separated to boards with varying topics of interest.
In this work, we focus on the Politically Incorrect board (\dspol) as it exhibits a high degree of racism and hate speech~\cite{hine2017kek} and it is an influential actor on the Web's information ecosystem~\cite{zannettou2017web}.
To obtain data from \dspol posts we use the same crawling infrastructure as discussed in \cite{hine2017kek}, while for the images we use the methodology discussed in \cite{zannettou2018origins}. Specifically, we obtain posts and images posted between July 2016 and January 2018, hence acquiring 67,416,903 posts and 5,859,439 images.

\descr{Gab.}
Gab is a newly created social network, founded in August 2016, that explicitly welcomes banned users from other communities (e.g., Twitter).
It waves the flag of free speech and it has mild moderation; it allows everything except illegal pornography, posts that promote terrorist acts, and doxing
other users. 
To obtain data from Gab, we use the same methodology as described in \cite{zannettou2018gab} and \cite{zannettou2018origins} for posts and images, respectively.
Overall, we obtain 35,528,320 posts and 1,125,154 images posted between August 2016 and January 2018.

\descr{Ethical Considerations.} During this work, we only collect publicly available data posted on \dspol and Gab.
We make no attempt to de-anonymize users. Overall, we follow best ethical practices as documented in~\cite{rivers2014ethical}.

\begin{table}[]
\centering
\resizebox{0.55\columnwidth}{!}{%
\begin{tabular}{@{}crr@{}}
\toprule
\multicolumn{1}{l}{\textbf{Platform}} & \multicolumn{1}{c}{\textbf{/pol/}} & \multicolumn{1}{c}{\textbf{Gab}} \\ \midrule
\textbf{\# of posts} & 67,416,903 & 35,528,320 \\
\textbf{\# of images} & 5,859,439 & 1,125,154 \\ \bottomrule
\end{tabular}%
}
\caption{Overview of our datasets. We report the number of posts and images from /pol/ and Gab.}
\label{tbl:datasets}
\end{table}

\section{Results}
In this section, we present our temporal analysis that shows the use of racial slurs over time on Gab and \dspol, our text-based analysis that leverages word2vec embeddings~\cite{mikolov2013efficient} to understand the use of language with respect to ethnic slurs, and our memetic analysis that focuses on the propagation of the antisemitic Happy Merchant meme. 
Finally, we present our influence estimation findings that shed light on the influence that Web communities have on each other when considering the spread of antisemitic memes.

\begin{table}[]
\centering
\resizebox{\columnwidth}{!}{%
\begin{tabular}{@{}lrrrrrr@{}}
\toprule
\multicolumn{4}{c}{\textbf{/pol/}} & \multicolumn{3}{c}{\textbf{Gab}} \\ \midrule
\multicolumn{1}{c}{\textbf{Term}} & \multicolumn{1}{c}{\textbf{\#posts (\%)}} & \multicolumn{1}{c}{\textbf{Rank}} & \multicolumn{1}{c|}{\textbf{\begin{tabular}[c]{@{}c@{}}Ratio\\ Increase\end{tabular}}} & \multicolumn{1}{c}{\textbf{\#posts (\%)}} & \multicolumn{1}{c}{\textbf{Rank}} & \multicolumn{1}{c}{\textbf{\begin{tabular}[c]{@{}c@{}}Ratio\\ Increase\end{tabular}}} \\ \midrule
\textbf{``jew''} & 1,993,432 (3.0\%) & 13 & \multicolumn{1}{r|}{1.64} & 763,329 (2.0\%) & 19 & 16.44 \\
\textbf{``kike''} & 562.983 (0.8\%) & 147 & \multicolumn{1}{r|}{2.67} & 86,395 (0.2\%) & 628 & 61.20 \\
\textbf{``white''} & 2,883,882 (4.3\%) & 3 & \multicolumn{1}{r|}{1.25} & 1,336,756 (3.8\%) & 9 & 15.92 \\
\textbf{``black''} & 1,320,213 (1.9\%) & 22 & \multicolumn{1}{r|}{0.89} & 600,000 (1.6\%) & 49 & 7.20 \\
\textbf{``nigger''} & 1,763,762 (2.6\%) & 16 & \multicolumn{1}{r|}{1.28} & 133,987 (0.4\%) & 258 & 36.88 \\ \midrule
\textbf{Total} & 67,416,903(100\%) & - & \multicolumn{1}{r|}{0.95} & 35,528,320(100\%) & - & 8.14 \\ \bottomrule
\end{tabular}
}
\caption{Number of posts, and their respective percentage in the dataset, for the terms``jew," ``kike,'' ``white,'' ``black,'' and ``nigger.'' We also report the rank of each term for each dataset (i.e., popularity in terms of count of appearance) and the ratio of increase between the start and the end of our datasets.}
\label{tbl:datasets_terms}
\end{table}

\begin{table}[t!]
\centering
\resizebox{0.3\columnwidth}{!}{%
\begin{tabular}{@{}lrr@{}}
\toprule
\textbf{Word}      & \multicolumn{1}{l}{\textbf{/pol/}} & \multicolumn{1}{l}{\textbf{Gab}} \\ \midrule
\textbf{``jew''}   & 42\%                               & 42\%                              \\
\textbf{``white''} & 33\%                               & 27\%                              \\
\textbf{``black''} & 43\%                               & 28\%                              \\ \bottomrule
\end{tabular}%
}
\caption{Percentage of hateful posts from random samples of 100 posts that include the words ``jew,'' ``white,'' and ``black.''}
\label{tbl:word_annotations}
\end{table}

\begin{figure}[t!]
\centering
\subfigure[/pol/]{\includegraphics[width=0.8\columnwidth]{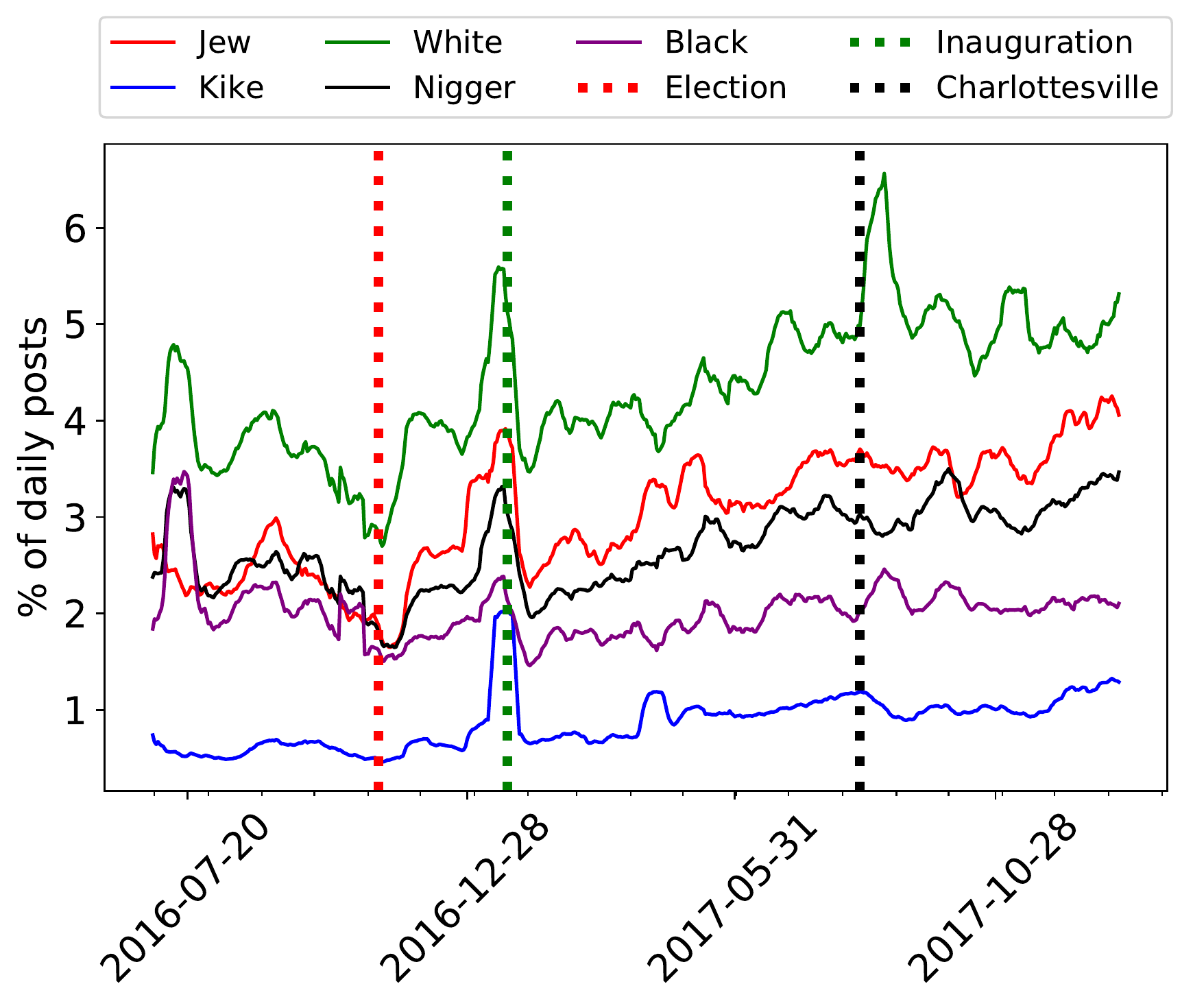}\label{fig:temporal_slurs_pol}}
\subfigure[Gab]{\includegraphics[width=0.8\columnwidth]{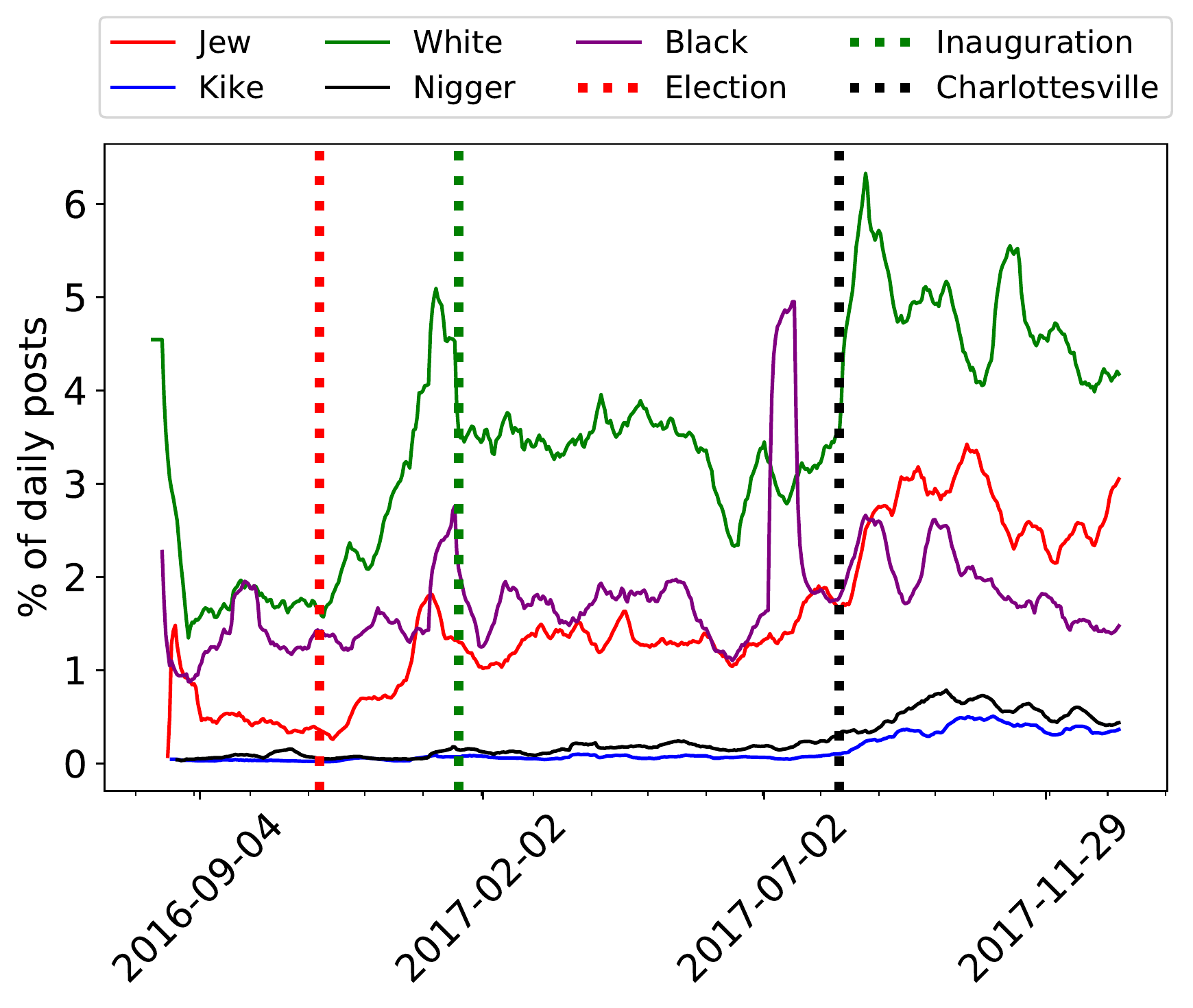}\label{fig:temporal_slurs_gab}}
\caption{Use of ethnic racial terms and slurs over time on /pol/ and Gab. Note that the vertical lines that show the three real-world events are indicative and are not obtained via rigorous time series analysis. The figure is best viewed in color.}
\label{fig:temporal_slurs}
\end{figure}

\begin{figure}[t!]
\centering
\subfigure[``jew'']{\includegraphics[width=0.9\columnwidth]{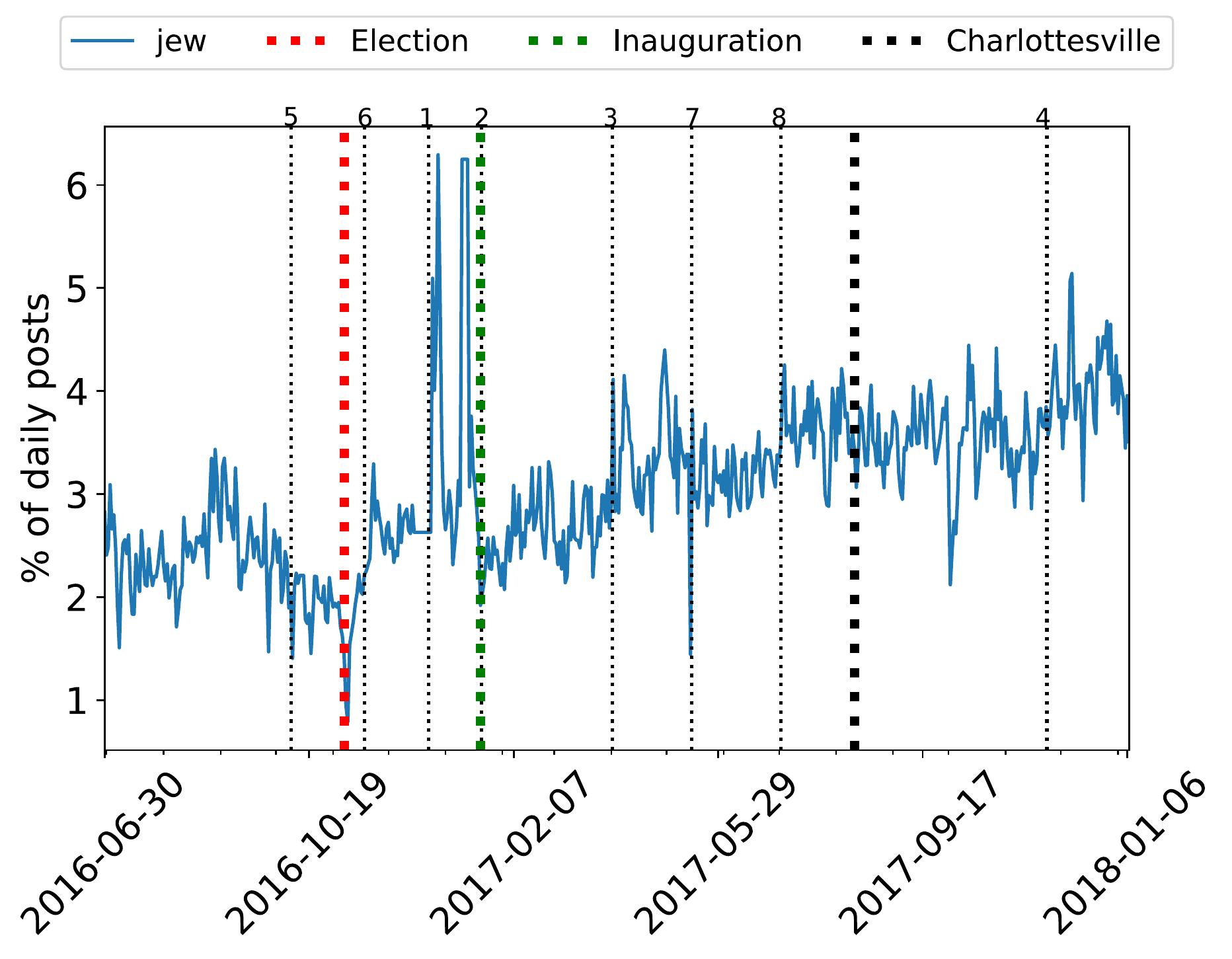}\label{fig:traces_jew_pol}}
\subfigure[``white'']{\includegraphics[width=0.9\columnwidth]{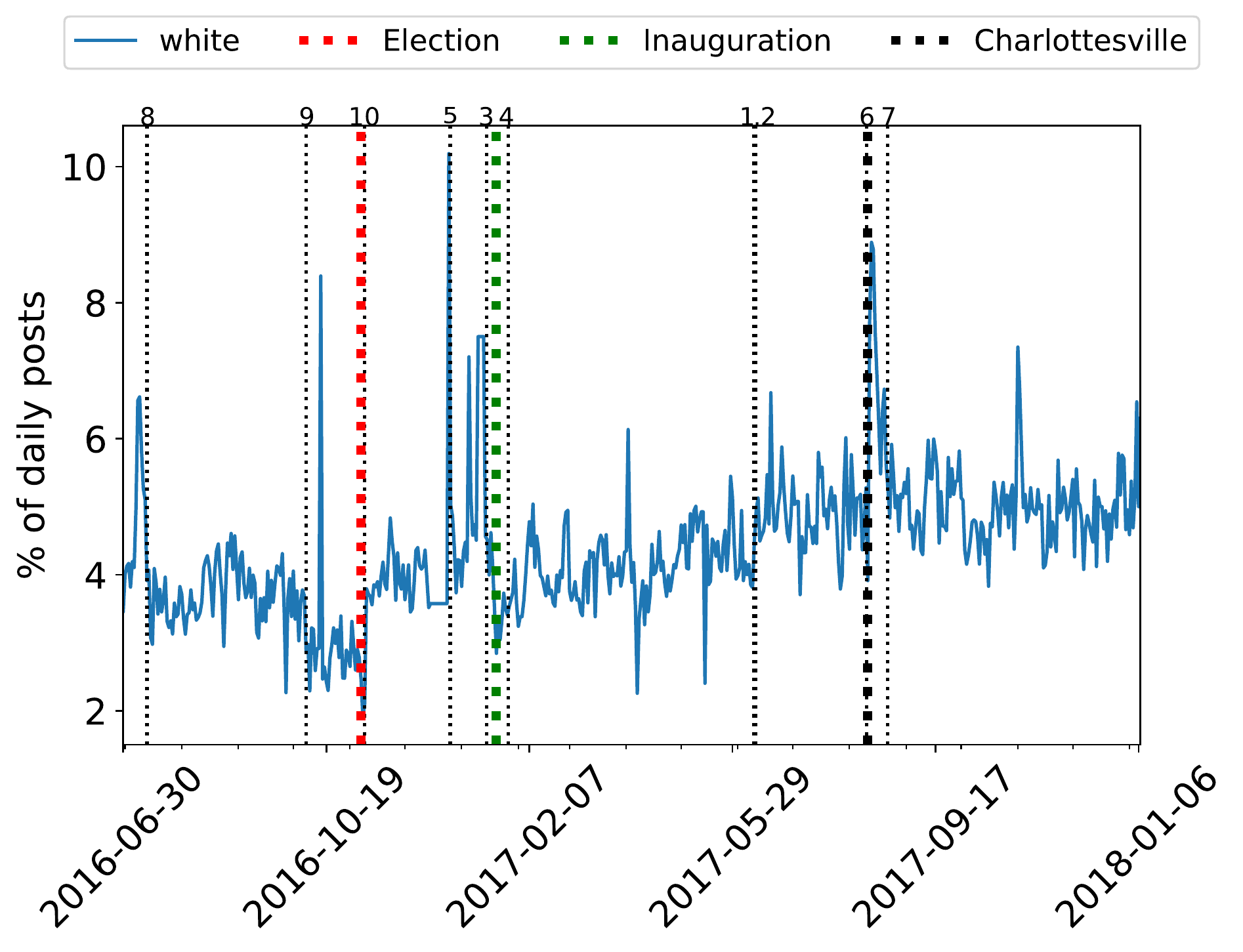}\label{fig:traces_white_pol}}
\caption{Percentage of daily posts per day for the terms ``jew'' and ``white'' on /pol/. We also report the detected changepoints (see Tables~\ref{tbl:jew_changepoints} and \ref{tbl:white_changepoints}, respectively, for the meaning of each changepoint). }
\label{fig:traces_pol}
\end{figure}

\begin{table*}[t]
\centering
\resizebox{\textwidth}{!}{%
\begin{tabular}{|c|l|l|}
\hline
\multicolumn{1}{|c|}{\textbf{Rank}} & \multicolumn{1}{c|}{\textbf{Date}} & \multicolumn{1}{c|}{\textbf{Events}} \\ \hline
1 & 2016-12-25 & \hspace{-0.55mm}\begin{tabular}[c]{@{}l@{}}
 2016-12-19: ISIS truck attack in Berlin Germany~\cite{berlin2017}.\end{tabular} \\ \hline
2 & 2017-01-17 & \hspace{-0.55mm}\begin{tabular}[c]{@{}l@{}}2017-01-17: Presidential inauguration of  Donald Trump~\cite{inaug2017}.\\ 2017-01-17: Benjamin Netanyahu attacks the latest peace-conference by calling it ``useless''~\cite{buet2017}.\end{tabular} \\ \hline
3 & 2017-04-02 & \hspace{-0.55mm}\begin{tabular}[c]{@{}l@{}}2017-04-05: President Trump removes Steve Bannon from his position on the National Security Council~\cite{bannonnsc2017}.\\ 2017-04-06: President Trump orders a strike on the Shayrat Air Base in Homs, Syria, using 59 Tomahawk cruise missiles~\cite{strike2017}.\end{tabular} \\ \hline
4 & 2017-11-26 & 2017-11-29: It is revealed that Jared Kushner has been interviewed by Robert Mueller's team in November~\cite{apuzzo2017}. \\ \hline
5 & 2016-10-08 & \hspace{-0.55mm}\begin{tabular}[c]{@{}l@{}}2016-10-09: Second presidential debate~\cite{debate2_2017}. \\ 2016-10-09: A shooting takes place in Jerusalem that kills a police officer and two innocent people, wounding several others~\cite{bbcgunman2016}.\end{tabular} \\ \hline
6 & 2016-11-20 & 2016-11-19: Swastikas, Trump Graffiti appear in Beastie Boys’ Adam Yauch Memorial Park in Brooklyn~\cite{chen2016}. \\ \hline
7 & 2017-05-16 & \hspace{-0.55mm}\begin{tabular}[c]{@{}l@{}}2017-05-16: Donald Trump admits that he shared classified information with Russian envoys~\cite{parker2017}.\\ 2017-05-16: U.S. intelligence warns Israel to withhold information from Trump, due to fears that it could fall into Russians or Iranians~\cite{moore2017}.\end{tabular} \\ \hline
8 & 2017-07-02 & \hspace{-0.55mm}\begin{tabular}[c]{@{}l@{}}2017-06-25: The Supreme Court reinstates President Trump's travel ban~\cite{scban2017}. \\ 2017-06-29: President Trump's partial travel ban comes into effect~\cite{baneffect2017}.\end{tabular} \\ \hline
\end{tabular}%
}
\caption{Dates that significant changepoint were detected in posts that contain the term ``jew'' on /pol/. We sort them according to their ``significance'' and we report corresponding real-world events that happened one week before/after of the changepoint date.}
\label{tbl:jew_changepoints}
\end{table*}

\begin{table*}[t]
\centering
\resizebox{\textwidth}{!}{%
\begin{tabular}{|c|l|l|}
\hline
\multicolumn{1}{|l|}{\textbf{Rank}} & \multicolumn{1}{c|}{\textbf{Date}} & \multicolumn{1}{c|}{\textbf{Events}} \\ \hline
\begin{tabular}[c]{@{}c@{}}1\\ 2\end{tabular} & \begin{tabular}[c]{@{}l@{}}2017-06-10\\ 2017-06-11\end{tabular} & \hspace{-0.55mm}\begin{tabular}[c]{@{}l@{}}2017-06-08: James Comey testifies about his conversations with Trump on whether he asked him to end investigations into Michael Flynn~\cite{politico2017}.\\ 2017-06-12: A federal court rejects Trump's appeal to stop the injunction against his travel ban~\cite{levine2017}.\\ 2017-06-13: The Senate Intelligence Committee interviews Jeff Sessions about potential Russian interference in the 2016 election~\cite{naughton2017}.\\ 2017-06-15: President Trump admits he is officially under investigation for obstruction of justice~\cite{shear2017}.\end{tabular} \\ \hline
3 & 2017-01-14 & 2017-01-17: Presidential inauguration of  Donald Trump~\cite{inaug2017}. \\ \hline
4 & 2017-01-24 & \hspace{-0.55mm}\begin{tabular}[c]{@{}l@{}}2017-01-23: Women's March protest~\cite{womens2017}.\\ 2017-01-25: President Trump formally issues executive order for construction of a wall on the United States - Mexico border~\cite{wall2017}.\end{tabular} \\ \hline
5 & 2016-12-25 & 2016-12-19: ISIS truck attack in Berlin Germany~\cite{berlin2017}.\\ \hline
6 & 2017-08-12 & \hspace{-0.55mm}\begin{tabular}[c]{@{}l@{}}2017-08-12: The ``Unite the Right'' rally takes place in Charlottesville, Virginia~\cite{unitetheright2017}.\\ 2017-08-13: President Trump condemns the violence from ``many sides'' at a far-right rally at Charlottesville~\cite{Lemire2017}.\end{tabular} \\ \hline
7 & 2017-08-21 & 2017-08-17: Steve Bannon resigns as Chief Strategist for the White House~\cite{bannon2017}. \\ \hline
8 & 2016-07-13 & \hspace{-0.55mm}\begin{tabular}[c]{@{}l@{}}
2016-07-08: Fatal shooting of 5 police officers in Dallas by Micha Xavier Johnson~\cite{dallas2016}.\\ 2016-07-14: Truck attack in Nice, France~\cite{nice2017}.\\ 2016-07-16: The 2016 Republican National Convention~\cite{convention2016}.\\
\end{tabular} \\ \hline
9 & 2016-10-08 & 2016-10-09: Second presidential debate~\cite{debate2_2017}. \\ \hline
10 & 2016-11-10 & 2016-11-08: Presidential election of Donald Trump~\cite{election2017}. \\ \hline
\end{tabular}%
}
\caption{Dates that significant changepoint were detected in posts that contain the term ``white'' on /pol/. We sort them according to their ``significance'' and we report corresponding real-world events that happened one week before/after of the changepoint date.}
\label{tbl:white_changepoints}
\end{table*}

\descr{Temporal Analysis.} Anecdotal evidence reports escalating racial and ethnic hate propaganda on fringe Web communities~\cite{thompson2018measure}.
To examine this, we study the prevalence of some terms related to ethnic slurs on \dspol and Gab, and how they evolve over time. 
We focus on five specific terms: ``jew,'' ``kike,'' ``white,'' ``black,'' and ``nigger.'' We limit our scope to these because while they are notorious for ethnic hate for many groups, these specific words ranked among the the most frequently used ethnic terms on both communities.
To extract posts for these terms, we first tokenize all the posts from \dspol and Gab, and then extract all posts that contain either of these terms. 
Note that we use the entire dataset without any further filters (e.g., we do not filter posts in other languages).
Table~\ref{tbl:datasets_terms} reports the overall number of posts that contain these terms in both Web communities, their rank in terms of raw number of appearances in our dataset, as well as the increase in the use of these terms between the beginning and end of our datasets.
For the latter, we note that although our computation of this ratio is in principle sensitive to large fluctuations at the ends of the dataset, Fig.~\ref{fig:temporal_slurs} do not display substantial fluctuations. Other methods, such as rolling averages, give comparable results. We study the effects of fluctuations systematically below.
Also, Fig.~\ref{fig:temporal_slurs} plots the use of these terms over time, binned by day, and averaged over a rolling window to smooth out small-scale fluctuations.
We annotate the figure with three real-world events, which are of great interest and are likely to cause change in activity in these fringe communities (according to our domain expertise). 
Namely, we annotate the graph with the 2016 US election day, the Presidential Inauguration, and the Charlottesville Rally.
We observe that terms like ``white'' and ``jew'' are extremely popular in both Web communities; 3rd and 13th respectively in \dspol, while in Gab they rank as the 9th and 19th most popular words, respectively.
We see a similar level of popularity for ethnic racial slurs like ``nigger'' and ``kike,'' especially on \dspol; they are the 16th and 147th most popular words in terms of raw counts.
Note that \dspol has a vocabulary 1.5x times larger than that of Gab (see Text Analysis below).
These findings highlight that both \dspol and Gab users habitually and increasingly engage in discussions about ethnicity and use targeted hate speech.

We also find an increasing trend in the use of most ethnic terms; the number of posts containing each of the terms except ``black'' increases, even when normalized for the increasing number of posts on the network overall.
Interestingly, among the terms we examine, we observe that the term ``kike'' shows the greatest increase in use for both \dspol and Gab, followed by ``jew'' on \dspol and ``nigger'' on Gab.
Also, it is worth noting that ethnic terms on Gab have a greater increase in the rate of use when compared to \dspol (cf. ratio of increase for \dspol and Gab in Table~\ref{tbl:datasets_terms}).
Furthermore, by looking at Fig.~\ref{fig:temporal_slurs} we find that by the end of our datasets, the term ``jew'' appears in 4.0\% of \dspol daily posts and 3.1\% of the Gab posts, while the term ``nigger'' appears in 3.4\% and 0.6\% of the daily posts on \dspol and Gab, respectively. %
The latter is particularly worrisome for anti-black hate, as by the end of our datasets the term ``nigger'' on \dspol overtakes the term ``black'' (3.4\% vs 1.9\% of all the daily posts).
Taken together, these findings highlight that most of these terms are increasingly popular within these communities, hence emphasizing the need to study the use of ethnic identity terms. 

To assess the extent that these terms are used in hateful/racist contexts we perform a small-scale manual annotation.
Specifically, we collect 100 random posts from \dspol and Gab for the words ``jew,'' ``white,'' and ``black'' and annotate them as hateful/racist or non-hateful/racist.
For each of these posts, an author of the paper inspects the post and, according to the tone and terminology used, labels it as being hateful/racist or not.
Note that we focus only on these three words, as the two other words (i.e., ``kike'' and ``nigger'') are highly offensive racial slurs, and therefore their use make the post immediately hateful/racist.
Table~\ref{tbl:word_annotations} report the percentage of hateful/racist posts for the random samples of posts obtained from \dspol and Gab.
We observe that these words are used in a hateful/racist context frequently: in our random sample more than 25\% of the posts that include one of the three words is hateful/racist.
We also find the least hateful/racist percentage for the term ``white'' mainly because it is used in several terms like ``White House'' or ``White Helmets'', while the same applies for the term ``black'' (to a lesser extent) and the ``Black Lives Matter'' movement.
Finally, we note a large hateful/racist percentage (42\%) for posts containing the term ``jew'', highlighting once again the emerging problem of antisemitism on both \dspol and Gab.

We note major fluctuations in the the use of ethnic terms over time, and one reasonable assumption is that these fluctuations happen due to real-world events.
To analyze the validity of this assumption, we use changepoint analysis, which provides us with
ranked changes in the mean and variance of time series behavior.
To perform the changepoint analysis, we use the PELT algorithm as described in~\cite{killick2012optimal}, and first applied to Gab timeseries data in~\cite{zannettou2018gab}.
We model each timeseries as a set of samples drawn from a normal distribution with mean and variance that are free to change at discrete times. 
We expect from the central limit theorem that for networks with large numbers of posts and actors, that this is a reasonable model.
The algorithm then fits a robust timeseries model to the data by finding the configuration of changepoints which maximize the likelihood of the observed data, subject to a penalty for the proliferation of changepoints.
The PELT algorithm thus returns the unique, exact best fit to the observed timeseries data.
Subject to the assumptions mentioned above, we are thus confident that the changepoints represent a meaningful aspect of the data.
We run the algorithm with a decreasing set of penalty amplitudes. 
We keep track of the largest penalty amplitude at which each changepoint first appears. 
This gives us a ranking of the changepoints in order of their ``significance.''

To identify real-world events that likely correspond to the detected changepoints, we manually inspect real-world events that are reported via the Wikipedia ``Current Events'' Portal\footnote{\url{https://en.wikipedia.org/wiki/Portal:Current_events}} and happened one week before/after of the changepoint date. 
The portal provides real-world events that happen across the world for each day.
To select the events, we use our domain expertise to identify the real-world events that are likely to be discussed by users on 4chan and Gab, hence they are the most likely events that caused the statistically significant change in the time series.

In \dspol, our analysis reveals several changepoints with temporal proximity to real-world political events for the use of both ``jew'' (see Fig.~\ref{fig:traces_jew_pol} and  Table~\ref{tbl:jew_changepoints}) and ``white'' (see Fig.~\ref{fig:traces_white_pol} and Table~\ref{tbl:white_changepoints}).
For usage in the term ``jew,'' major world events in Israel and the Middle East correspond to several changepoints, including the U.S. missile attack against Syrian airbases in 2017, and terror attacks in Jerusalem.  
Events involving Donald Trump like the resignation of Steve Bannon from the National Security Council, the 2017 ``travel ban'' (i.e., Executive Order 13769), and the presidential inauguration occur within proximity to several notable changepoints for usage of ``jew'' as well. 
For ``white,'' we find that changepoints correspond closely to events related to Donald Trump, including the election, inauguration, presidential debates, as well as major revelations in the ongoing investigation into Russian interference in the presidential election.  
Additionally, several changepoints correspond to major terror attacks by ISIS in Europe, including vehicle attacks in Berlin and Nice, as well as news related to the 2017 ``travel ban'' (i.e., Executive Order 13769).  
In the case of ``white,'' the relationship between online usage and real-world behavior is best illustrated by the Charlottesville ``Unite the Right'' rally, which marks the global maximum in our dataset for the use of the term on both \dspol and Gab (see Fig.~\ref{fig:temporal_slurs}). 
For Gab, we find that changepoints in these time series reflect similar kinds of news events to those in \dspol, both for ``jew'' and ``white'' (we omit the Figures and Tables due to space constraints). %
These findings provide evidence that discussion of ethnic identity on fringe communities increases with political events and real-world extremist actions.
The implications of this relationship are worrying, as others have shown that ethnic hate expressed on social media influences real-life hate crimes~\cite{muller2018making,muller2017fanning}.

\begin{table}[]
\centering
\resizebox{\columnwidth}{!}{%
\begin{tabular}{lrlrlrlr}
\hline
\multicolumn{4}{c}{\textbf{/pol/}} & \multicolumn{4}{c}{\textbf{Gab}} \\ \hline
\textbf{Word} & \multicolumn{1}{l}{\textbf{\begin{tabular}[c]{@{}l@{}}Similarity\end{tabular}}} & \textbf{Word} & \multicolumn{1}{r|}{\textbf{Probability}} & \textbf{Word} & \multicolumn{1}{l}{\textbf{\begin{tabular}[c]{@{}l@{}}Similarity\end{tabular}}} & \textbf{Word} & \multicolumn{1}{l}{\textbf{Probability}} \\ \hline
(((jew))) & 0.802 & ashkenazi & \multicolumn{1}{r|}{0.269} & jewish & 0.807 & jew & 0.770 \\
jewish & 0.797 & jew & \multicolumn{1}{r|}{0.196} & kike & 0.777 & jewish & 0.089 \\
kike & 0.776 & jewish & \multicolumn{1}{r|}{0.143} & gentil & 0.776 & gentil & 0.044 \\
zionist & 0.723 & outjew & \multicolumn{1}{r|}{0.077} & goyim & 0.756 & shabbo & 0.014 \\
goyim & 0.701 & sephard & \multicolumn{1}{r|}{0.071} & zionist & 0.735 & ashkenazi & 0.013 \\
gentil & 0.696 & gentil & \multicolumn{1}{r|}{0.026} & juden & 0.714 & goyim & 0.005 \\
jewri & 0.683 & zionist & \multicolumn{1}{r|}{0.025} & (((jew))) & 0.695 & kike & 0.005 \\
zionism & 0.681 & hasid & \multicolumn{1}{r|}{0.024} & khazar & 0.688 & zionist & 0.005 \\
juden & 0.665 & talmud & \multicolumn{1}{r|}{0.010} & jewri & 0.681 & rabbi & 0.004 \\
heeb & 0.663 & mizrahi & \multicolumn{1}{r|}{0.006} & yid & 0.679 & talmud & 0.003 \\ \hline
\end{tabular}%
}
\caption{Top ten similar words to the term ``jew'' and their respective cosine similarity. We also report the top ten words generated by providing as a context term the word ``jew'' and their respective probabilities on \dspol and Gab. }
\label{tbl:jew_w2v}
\end{table}

\descr{Text Analysis.}
We hypothesize that ethnic terms (e.g., ``jew'' and ``white'') are strongly linked to antisemitic and white supremacist sentiments.
To test this, we use word2vec, a two-layer neural network that generate word representations as embedded vectors~\cite{mikolov2013efficient}.
Specifically, a word2vec model takes as an input a large corpus of text and generates a multi-dimensional vector space where each word is mapped to a vector in the space (also called an embedding).
The vectors are generated in such way that words that share similar contexts tend to have nearly parallel vectors in the multi-dimensional vector space.
Given a context (list of words appearing in a single block of text), a trained word2vec model also gives the probability that each other word will appear in that context. By analyzing both these probabilities and the word vectors themselves, we are able to map the usage of various terms in our corpus.

We train two word2vec models; one for the \dspol dataset and one for the Gab dataset. 
First, as a pre-processing step, we remove stop words (such as ``and,'' ``like,'' etc.), punctuation, and we stem every word. 
Then, using the words of each post we train our word2vec models with a context window equal to 7 (defines the maximum distance between the current and the predicted words during the generation of the word vectors). 
We elect to slightly increase the context window from the default 5 to 7, since posts on \dspol tend to be longer when compared to other platforms like Twitter.
Also, we consider only words that appear at least 500 times in each corpus, hence creating a vocabulary of 31,337 and 20,115 stemmed words for \dspol and Gab, respectively.
Next, we use the generated word embeddings to gain a deeper understanding of the \emph{context} in which certain terms are used. 
We measure the ``closeness'' of two terms ($i$ and $j$) by generating their vectors from the word2vec models ($h_i$ and $h_j$) and calculating their cosine similarity ($\cos\theta(h_1, h_2)$).
Furthermore, we use the trained models to predict a set of candidate words that are likely to appear in the context of a given term.

We first look at the term ``jew.''
Table~\ref{tbl:jew_w2v} reports the top ten most similar words to the term ``jew'' along with their cosine similarity, as well as the top ten candidate words and their respective probability. %
By looking to the most similar words, we observe that on \dspol ``(((jew)))'' is the most similar term ($\cos\theta = 0.80$), while on Gab is the 7th most similar term ($\cos\theta = 0.69$).
The triple parentheses is a widely used, antisemitic symbol that calls attention to supposed secret Jewish involvement and conspiracy~\cite{schama2018semitism}.
Slurs like ``kike,'' which is historically associated with
general ethnic disgust, rank similarly ($\cos\theta = 0.77$ on both \dspol and Gab).
This suggests that on both Web communities, the term ``jew'' itself is closely related to classical antisemitic contexts.
When digging deeper, we note that ``goyim'' is the 5th and 4th most similar term to ``jew,'' in \dspol and Gab, respectively. 
``Goyim'' is the plural of ``goy,'' and while its original meaning is just ``non-jews,'' modern usage tends to have a derogatory nature~\cite{goyim_wiki}.
On fringe Web communities it is used to emphasize the ``struggle'' against Jewish conspiracy by preemptively assigning Jewish hostility to non-Jews as in ``The Goyim Know'' meme~\cite{the_goyim_know_meme}. 
It is also commonly used in a dismissive manner toward community members; a typical attacker will accuse a user he disagrees with of being a ``good goy,''~\cite{good_goy} a meme implying obedience to a supposed Jewish elite conspiracy.

When looking at the set of candidate words, given the term ``jew,'' we find the candidate word ``ashkenazi'' (most likely on \dspol and 5th most likely on Gab), which refers to a specific subset of the Jewish community.
Interestingly, we note that the term ``jew'' exists in the set of most likely words for both communities, hence indicating that \dspol and Gab users abuse the term ``jew'' by posting messages that include the term ``jew'' multiple times in the same sentence.
We also note that this has a higher probability of happening on Gab rather than \dspol (cf. probabilities for candidate word ``jew'' in Table~\ref{tbl:jew_w2v}).

\begin{figure}[t]
\centering
\resizebox{0.7\columnwidth}{!}{
\includegraphics[width=\columnwidth]{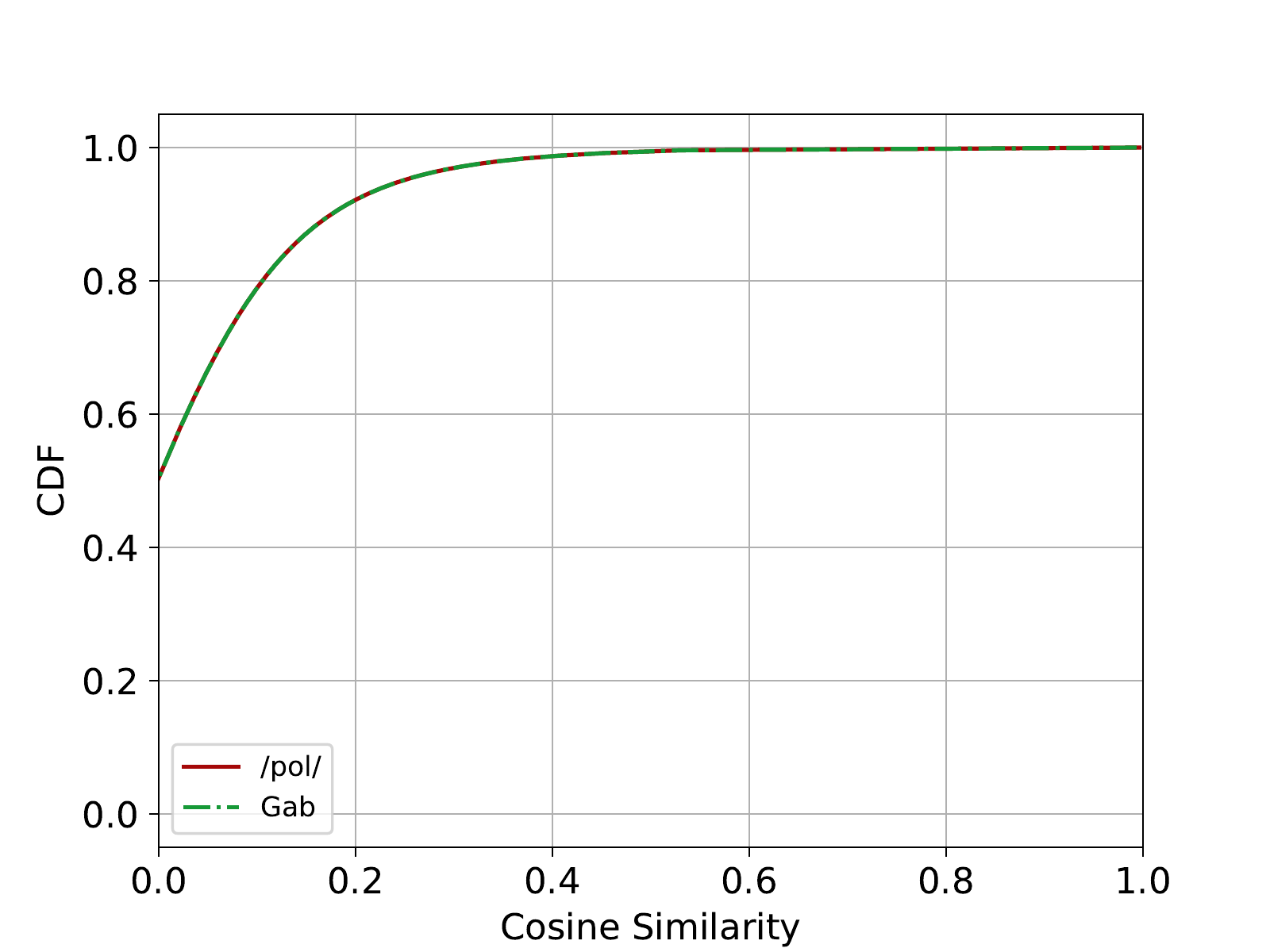}
}
\caption{CDF of the cosine similarities for all the pairs of words in the trained word2vec models. }
\label{fig:cdf_cosine_distances_complete}
\end{figure}

\begin{figure*}[t!]
\centering
\includegraphics[width=\textwidth]{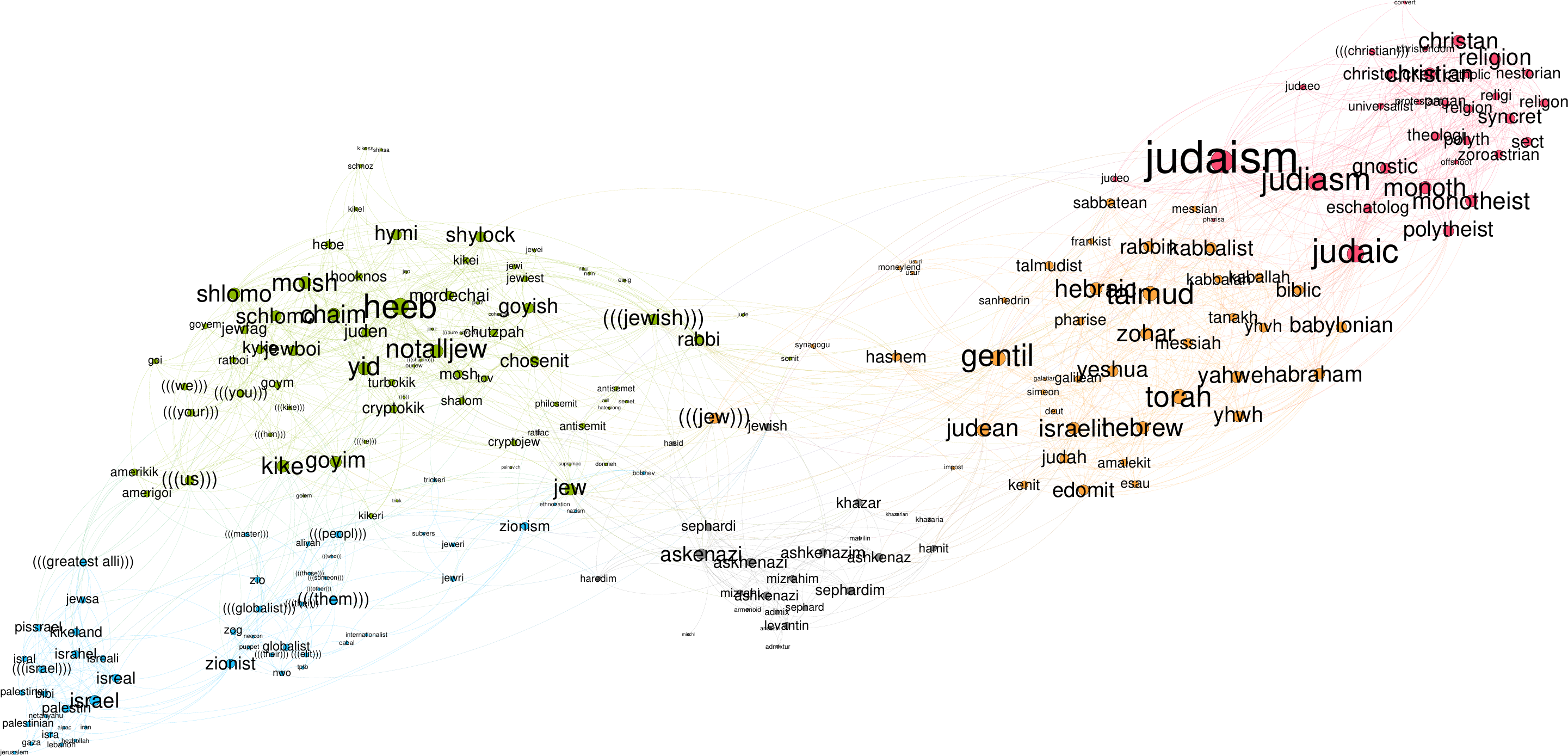}
\caption{Graph representation of the words associated with ``jew'' on \dspol. We extract the graph by finding the most similar words, and then we take the 2-hop ego network around ``jew.’’ In this graph the size of a node is proportional to its degree; the color of a node is based on the community it is a member of; and the entire graph is visualized using a layout algorithm that takes edge weights into account (i.e.,  nodes with similar words will be closer in the visualization). Note that the figure is best viewed in color.}
\label{fig:jew_graph}
\end{figure*}

To better show the connections between words similar to ``jew,'' Fig.~\ref{fig:jew_graph} demonstrates the words associated with ``jew'' on \dspol as a graph (we omit the same graph for Gab due to space constraints), where nodes are words obtained from the word2vec model, and the edges are weighted by the cosine similarities between the words (obtained from the trained word2vec models).
The graph visualizes the two-hop ego network~\cite{ego_network} from the word ``jew,'' which includes all the nodes that are either directly connected or connected through an intermediate node to the ``jew'' node. 
We consider two nodes to be connected if their corresponding word vectors have a cosine similarity that is greater or equal to a pre-defined threshold.
To select this threshold, we inspect the CDF of the cosine similarities between all the pair of words that exist in the trained word2vec models (we omit the figure due to space constraints).
We elect to set this threshold to 0.6, which corresponds to keeping only 0.2\% of all possible connections (cosine similarities).
We argue that this threshold is reasonable as all the pairwise pairs of cosine similarities between the words is an extremely large number.
To identify the structure and communities in our graph, we run the community detection heuristic presented in~\cite{blondel2008fast}, and we paint each community with a different color.
Finally, the graph is layed out with the ForceAtlas2 algorithm~\cite{jacomy2014forceatlas2}, which takes into account the weight of the edges when laying out the nodes in the 2-dimensional space.

This visualization reveals the existence
of historically salient antisemitic terms, as well as newly invented slurs, as the most prominent associations to the word ``jew.''
We also note communities forming distinct themes.
Keeping in mind that proximity in the visualization implies contextual similarity, we note two close, but distinct communities of words which portray Jews as a morally corrupt ethnicity on the one hand (green nodes), and as powerful geopolitical conspirators on the other (blue). 
Notably the blue community connects canards of Jewish political power to anti-Israel and anti-Zionist slurs.
The three, more distant communities document \dspol's interest in three topics: The obscure details of ethnic Jewish identity (grey),  Kabbalistic and cryptic Jewish lore (orange), and religious, or theological topics (pink).

We next examine the use of the term ``white.'' We hypothesize that this term is closely tied to ethnic nationalism.
To provide insight for how ``white'' is used on \dspol and Gab, we use the same analysis as described above for  the term ``jew.''
Table~\ref{tbl:white_w2v} shows the top ten similar words to ``white'' and the top ten most likely words to appear in the context of ``white.'' 
When looking at the most similar terms, we note the existence of ``huwhite'' ($\cos\theta=0.78$ on \dspol and $\cos\theta=0.70$ on Gab), a pronunciation of ``white'' popularized by the YouTube videos of white supremacist, Jared Taylor~\cite{taylor2017}. 
``Huwhite'' is a particularly interesting example of how the alt-right adopts certain language, even language that is seemingly derogatory towards themselves, in an effort to further their ideological goals.
We also note the existence of other terms referring to ethnicity, such the terms ``black'' ($\cos\theta=0.77$ on \dspol and $\cos\theta=0.71$ on Gab), ``whiteeuropean'' ($\cos\theta=0.64$ on \dspol), and ``caucasian'' ($\cos\theta=0.64$ on Gab).
Interestingly, we again note the presence of the triple parenthesis ``(((white)))'' term on \dspol ($\cos\theta=0.75$), which refers to Jews who conspire to disguise themselves as white. 
When looking at the most likely candidate words, we find that on \dspol the term ``white'' is linked with ``supremacist,'' ``supremacy,'' and other ethnic nationalism terms.
The same applies on Gab with greater intensity as the word ``supremacist'' has a substantially larger probability when compared to \dspol.

\begin{figure}[t!]
\centering
\includegraphics[width=0.9\columnwidth]{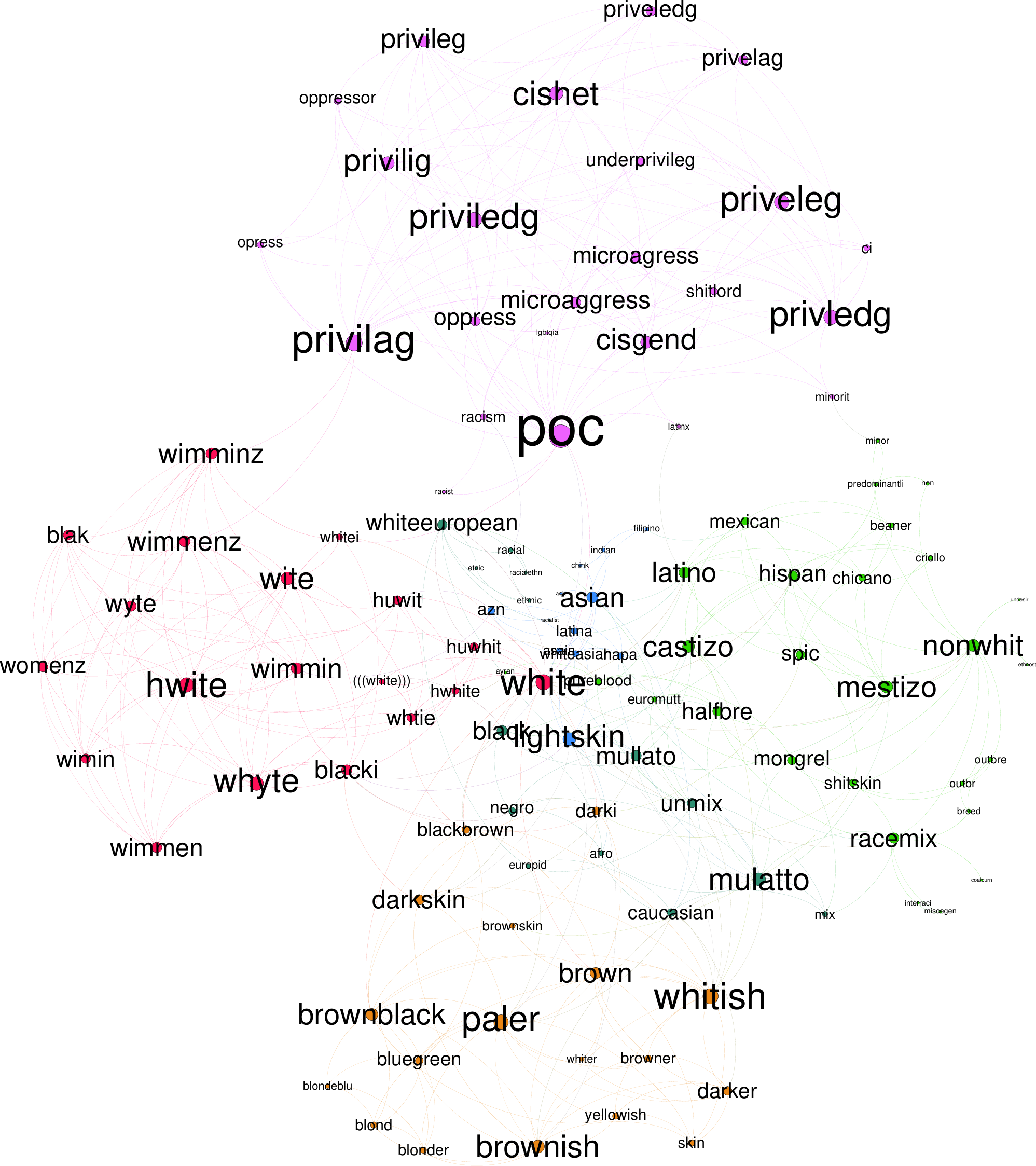}
\caption{Graph representation of the words associated with ``white'' on \dspol. Note that the figure is best viewed in color.}
\label{fig:white_graph}
\end{figure}

To provide more insight into the contexts and use of ``white'' on \dspol we show its most similar terms and their nearest associations in Fig.~\ref{fig:white_graph} (using the same approach as for ``jew'' in Fig.~\ref{fig:jew_graph}, we omit the same graph for Gab due to space constraints).
We find six different communities that evidence identity politics alongside themes of racial purity, miscegenation, and political correctness.
These communities correspond to distinct ethnic and gender themes, like Hispanics (green), Blacks (orange), Asians (blue), and women (red).
The final two communities relate to concerns about race-mixing (teal) and a prominent pink cluster that intriguingly, references terms related to left-wing political correctness~\cite{burch2018microaggression}, such as microagression and privilege (violet).

Note that we made the same analysis for the rest of the words that we study (i.e., ``kike,'' ``nigger,'' and ``black''), however, we omit the figures and analysis due to space constraints.

\begin{table}[]
\centering
\resizebox{\columnwidth}{!}{%
\begin{tabular}{@{}lrlrlrlr@{}}
\toprule
\multicolumn{4}{c}{\textbf{/pol/}} & \multicolumn{4}{c}{\textbf{Gab}} \\ \midrule
\textbf{Word} & \multicolumn{1}{l}{\textbf{\begin{tabular}[c]{@{}l@{}}Similarity\end{tabular}}} & \textbf{Word} & \multicolumn{1}{r|}{\textbf{Probability}} & \textbf{Word} & \multicolumn{1}{l}{\textbf{\begin{tabular}[c]{@{}l@{}}Similarity\end{tabular}}} & \textbf{Word} & \multicolumn{1}{l}{\textbf{Probability}} \\ \midrule
huwhit & 0.789 & supremacist & \multicolumn{1}{r|}{0.494} & black & 0.713 & supremacist & 0.827 \\
black & 0.771 & supremaci & \multicolumn{1}{r|}{0.452} & huwhit & 0.703 & supremaci & 0.147 \\
(((white))) & 0.754 & supremist & \multicolumn{1}{r|}{0.008} & nonwhit & 0.684 & genocid & 0.009 \\
nonwhit & 0.747 & male & \multicolumn{1}{r|}{0.003} & poc & 0.669 & helmet & 0.004 \\
huwit & 0.655 & race & \multicolumn{1}{r|}{0.002} & caucasian & 0.641 & nationalist & 0.003 \\
hwite & 0.655 & supremecist & \multicolumn{1}{r|}{0.002} & whitepeopl & 0.625 & hous & 0.003 \\
whiteeuropean & 0.644 & nationalist & \multicolumn{1}{r|}{0.002} & dispossess & 0.624 & privileg & $<0.001$ \\
hispan & 0.631 & genocid & \multicolumn{1}{r|}{0.002} & indigen & 0.602 & male & $<0.001$ \\
asian & 0.628 & non & \multicolumn{1}{r|}{0.001} & negroid & 0.599 & knight & $<0.001$ \\
brownblack & 0.627 & guilt & \multicolumn{1}{r|}{0.001} & racial & 0.595 & non & $<0.001$ \\ \bottomrule
\end{tabular}%
}
\caption{Top ten similar words to the term ``white'' and their respective cosine similarity. We also report the top ten words generated by providing as a context term the word ``white'' and their respective probabilities on \dspol and Gab. }
\label{tbl:white_w2v}
\end{table}

 \begin{figure}[t]
\centering
\subfigure[/pol/]{\includegraphics[width=0.75\columnwidth]{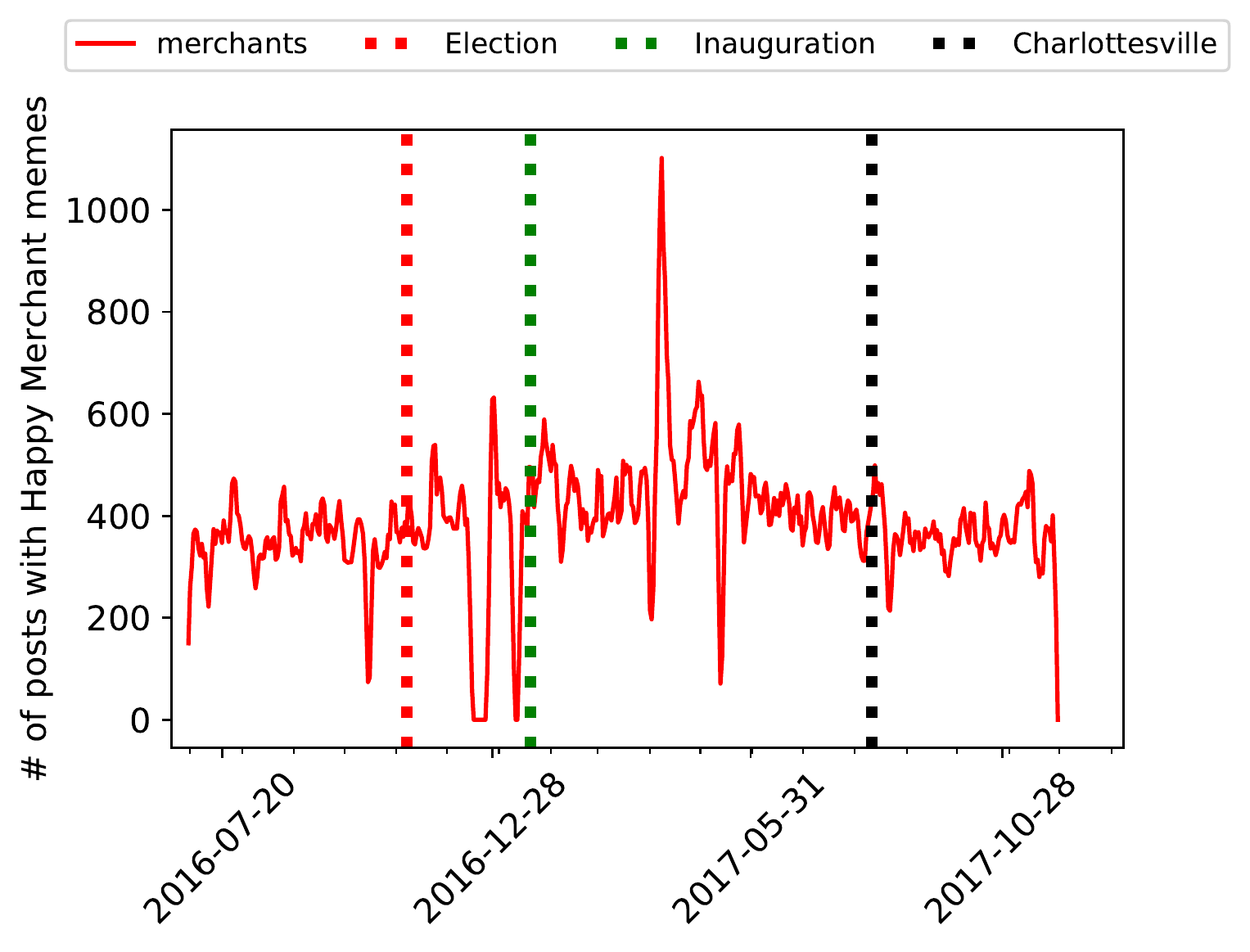}\label{fig:merchant_temporal_pol}}
\subfigure[Gab]{\includegraphics[width=0.75\columnwidth]{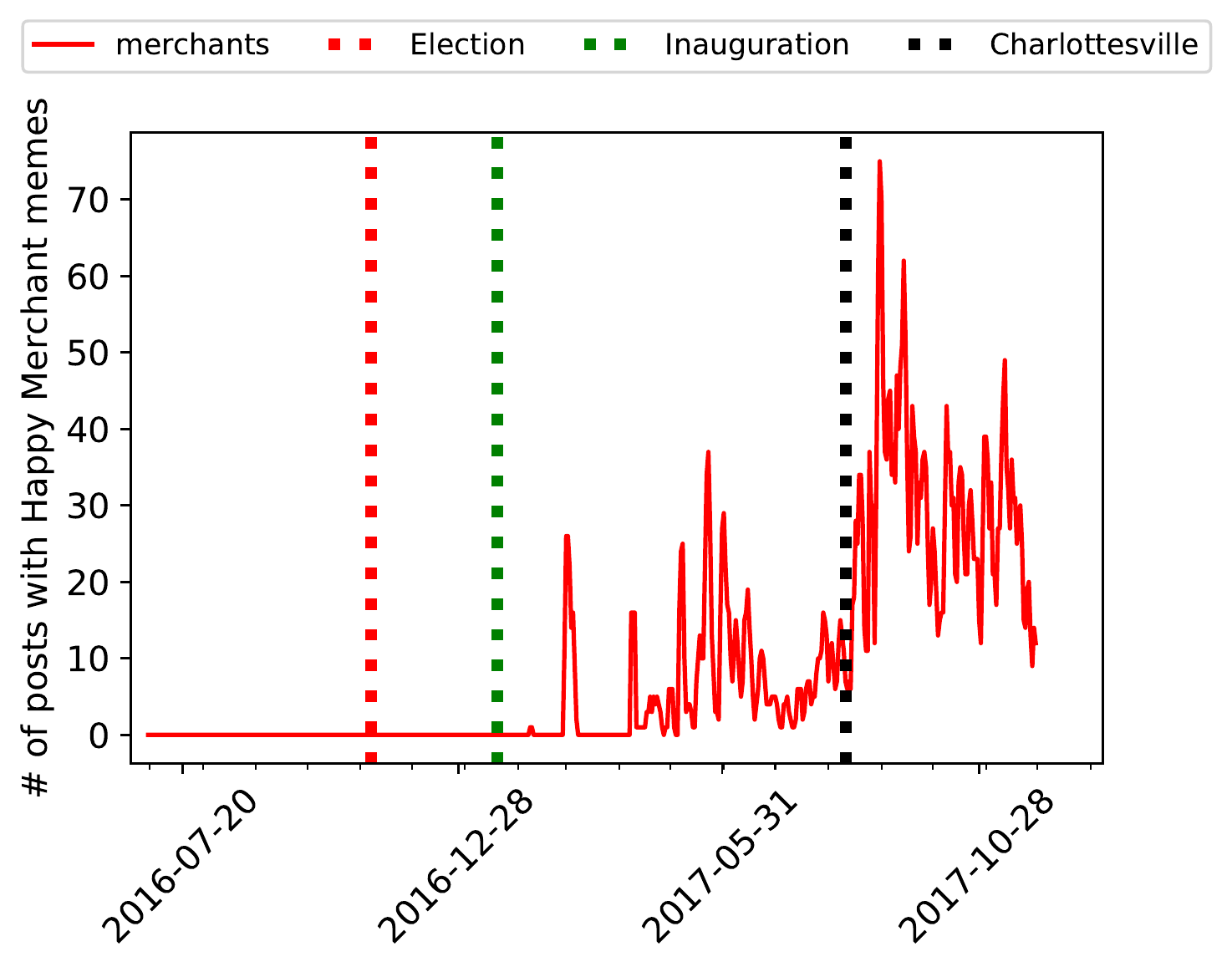}\label{fig:merchant_temporal_gab}}
\caption{Number of posts that contain images with the Happy Merchant meme on \dspol and Gab. Note that the vertical lines that show the three real-world events are indicative and are not obtained via our rigorous changepoint analysis.}
\label{fig:temporal_merchant}
\end{figure}

\begin{figure*}[t!]
\centering
\includegraphics[width=0.7\textwidth]{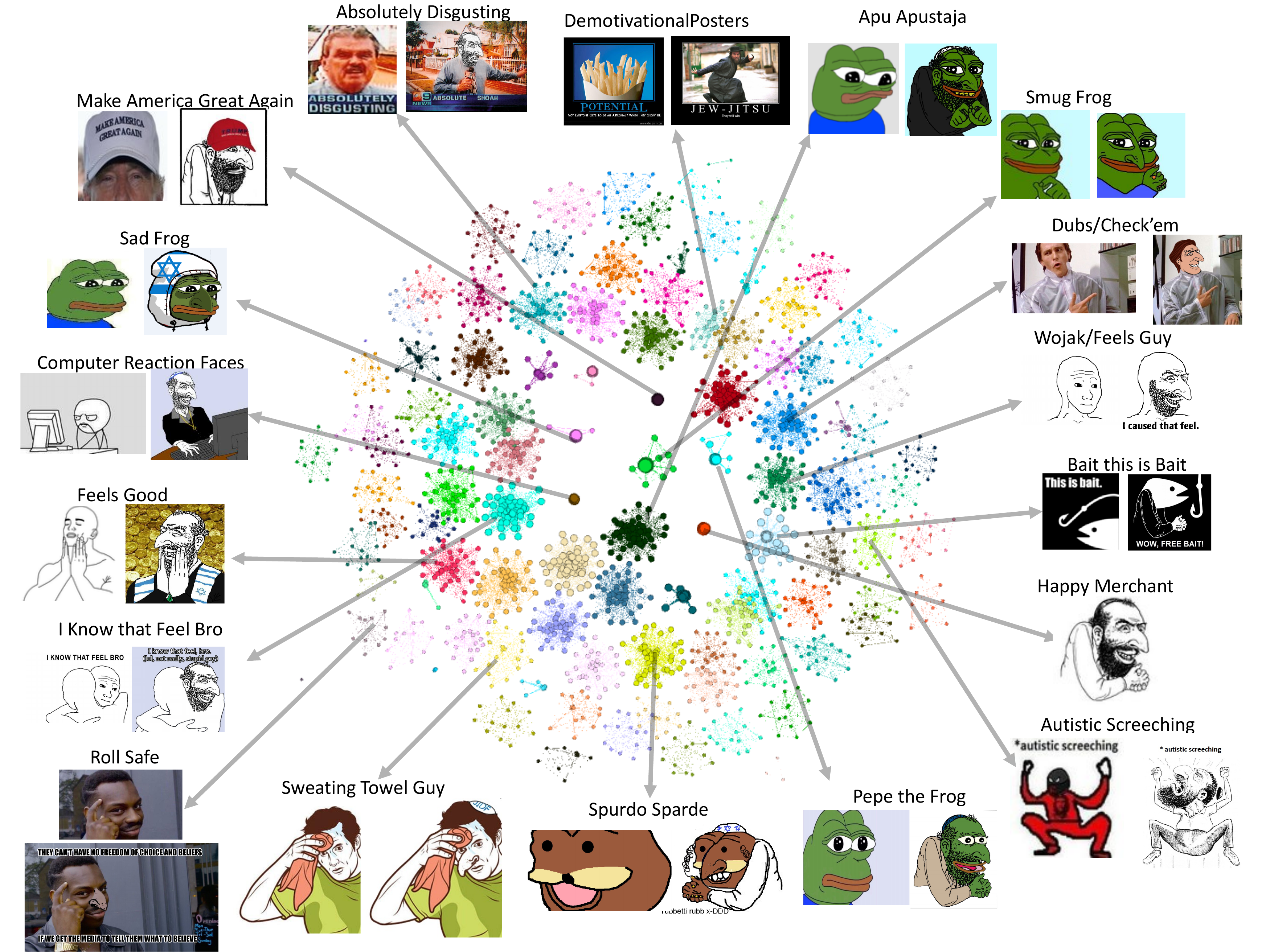}
\caption{Visualization of a subset of the obtained image clusters with a focus on the penetration of the Happy Merchant meme to other popular memes. The figure is inspired from~\cite{zannettou2018origins}.}
\label{fig:islands_annon_jews}
\end{figure*}

\descr{Meme Analysis.} In addition to hateful terms, memes also play a well documented role in the spread of propaganda and ethnic hate in Web communities~\cite{zannettou2018origins}.
To detail how memes spread and how different Web communities influence one another with memes, previous work~\cite{zannettou2018origins} established a pipeline that is able to track memes across multiple platforms. %
In a nutshell, the pipeline uses perceptual hashing~\cite{monga2006perceptual} and clustering techniques~\cite{ester1996density} to track and analyze the propagation of memes across multiple Web communities.
To achieve this, it relies on images obtained from the Know Your Meme (KYM) site~\cite{kym_site}, which is a comprehensive encyclopedia of memes.

In this work, we use this pipeline to study how antisemitic memes spread within and between these Web communities, and examine which communities are the most influential in their spread.
To do this, we additionally examine two mainstream Web communities, Twitter and Reddit, and compare their influence (with respect to memes) with \dspol and Gab.
For Twitter and Reddit, we use the dataset from~\cite{zannettou2018origins}, which includes all the posts from Reddit and Twitter, between July 2016 and July 2017, that include an image that is a meme as dictated by the KYM dataset and their processing pipeline. 
The final dataset consists of 581K tweets and 717K Reddit posts that include a meme.
In this work, we focus on the Happy Merchant meme (see Fig.~\ref{fig:merchant-example})~\cite{happy_merchant_meme}, which is an important hate-meme to study in this regard for several reasons. First, it represents an unambiguous instance of antisemitic hate, and second, it is extremely popular and diverse in \dspol and Gab~\cite{zannettou2018origins}.

We aim to assess the popularity and increase of use over time of the Happy Merchant meme on \dspol and Gab.
Fig.~\ref{fig:temporal_merchant} shows the number of posts that contain images with the Happy Merchant meme for every day of our \dspol and Gab dataset.
We further note that the numbers here represent a \emph{lower bound} on the number of Happy Merchant postings: the image processing pipeline is conservative and only labels clusters that are unambiguously the Happy Merchant; variations of other memes that incorporate the Happy Merchant are harder to assess.
\footnote{We refer readers to the extended version of the original paper~\cite{zannettou2018origins} for the assessment of the pipeline's performance.}
We observe that \dspol consistently shares antisemitic memes over time with a peak in activity on April 7, 2017, around the time that the USA launched a missile strike in a Syrian base~\cite{syrian_strike}. 
By manually examining a few posts including the Happy Merchant meme on this specific date, we find that 4chan users use this meme to express their belief that the Jews are ``behind this attack.''
On Gab we note a substantial and sudden
increase in posts containing Happy Merchant memes immediately after the Charlottesville rally. 
Our findings on Gab dramatically illustrate the implication that real-world eruptions of antisemitic behavior can catalyze the acceptability and popularity of antisemitic memes on other Web communities.%
Taken together, these findings highlight that both communities are exploited by users to disseminate racist content that is targeted towards the Jewish community. 

Another important step in examining the Happy Merchant meme is to explore how clusters of similar Happy Merchant memes relate to other meme clusters in our dataset.
One possibility is that Happy Merchants make-up a unique family of memes, which would suggest that they segregate in form and shape from other memes.
Given that many memes evolve from one another, a second possibility is that Happy Merchants ``infect'' other common memes.
This could serve, for instance, to make antisemitism more accessible and common.
To this end, we visualize in Fig.~\ref{fig:islands_annon_jews} a subset of the meme clusters, which we annotate using our KYM dataset, and a Happy Merchant version of each meme.
This visualization is inspired from~\cite{zannettou2018origins} and it demonstrates numerous instances of the Happy Merchant infecting well-known and popular memes.
Some examples include Pepe the Frog~\cite{pepe_frog_meme}, Roll Safe~\cite{roll_safe_meme}, Bait this is Bait~\cite{bait_meme}, and the Feels Good meme~\cite{feels_good_meme}.
This suggests that users generate antisemitic variants on recognizable and popular memes.

\begin{table*}[t!]
\centering
\resizebox{0.7\textwidth}{!}{%
\begin{tabular}{@{}lrrrrrrr@{}}
\toprule
\textbf{}                    & \multicolumn{1}{l}{\textbf{/pol/}} & \multicolumn{1}{l}{\textbf{Reddit}} & \multicolumn{1}{l}{\textbf{Twitter}} & \multicolumn{1}{l}{\textbf{Gab}} & \multicolumn{1}{l}{\textbf{T\_D}} & \multicolumn{1}{l}{\textbf{Total Events}} & \multicolumn{1}{l}{\textbf{\# of clusters}} \\ \midrule
\textbf{Happy Merchant Meme} & 43,419                             & 1,443                               & 1,269                                & 376                              & \multicolumn{1}{r|}{282}          & \multicolumn{1}{r|}{46,789}               & 133                                         \\
\textbf{Other Memes}         & 1,530,821                          & 581,244                             & 717,752                              & 44,542                           & \multicolumn{1}{r|}{81,665}       & \multicolumn{1}{r|}{2,956,024}            & 12,391                                      \\ \bottomrule
\end{tabular}%
}
\caption{Events per Web community for the Happy Merchant and all the other memes.}
\label{tbl:hawkes_events}
\end{table*}

\descr{Influence Estimation.} While the growth and diversity of the Happy Merchant within fringe Web communities is a cause of significant concern, a critical question remains: How do we chart the influence of Web communities on one another in spreading the Happy Merchant?
We have, until this point, examined the expanse of antisemitism on individual, fringe Web communities.
Memes however, develop with the purpose to replicate and spread between different Web communities.
To examine the influence of meme spread between Web communities, we employ Hawkes processes~\cite{linderman2014,lindermanArxiv}, which can be exploited to measure the predicted, reciprocal influence that various Web communities have to each other.
Generally, a Hawkes model consists of $K$ processes, where a process is a sequence of events that happen with a particular probability distribution.
Colloquially, a process is analogous to a specific Web community where memes (i.e., events) are posted.
Each process has a rate of events, which defines expected frequency of events on a specific Web community 
(for example, five posts with Happy Merchant memes per hour).
An event on one process can cause \textit{impulses} on other processes, which increase their rates for a period of time.
An impulse is defined by a weight and a probability distribution. 
The former dictates the intensity of the impulse (i.e., how strong is the increase in the rate of a process), while the latter dictates how the effect of the impulse changes over time (typically it decays as time goes on).
For instance, a weight of 1.5 from process A to B, means that each event on A will cause, on average, an additional 1.5 events on B.

In this work, we use a separate Hawkes model for each cluster of images that we obtained when applying the pipeline reported in~\cite{zannettou2018origins}.
Each model consists of five processes; one for each of \dspol, \td, the rest of Reddit, Gab, and Twitter.
We elected to separate \td from the rest of Reddit, as it is an influential actor with respect to the dissemination of memes~\cite{zannettou2018origins}.
Next, we fit each model using Gibbs sampling as reported in~\cite{linderman2014,lindermanArxiv}.
This technique enable us to obtain, at a given time, the weights and probability distributions for each impulse that is active, hence allowing us to be confident that an event is caused because of a previous event on the same or on another process.
Table~\ref{tbl:hawkes_events} shows the number of events (i.e., appearance of a meme) for each community we study, for both the Happy Merchant meme and all the other memes.

\begin{figure}[t!]
\centering
\includegraphics[width=\columnwidth]{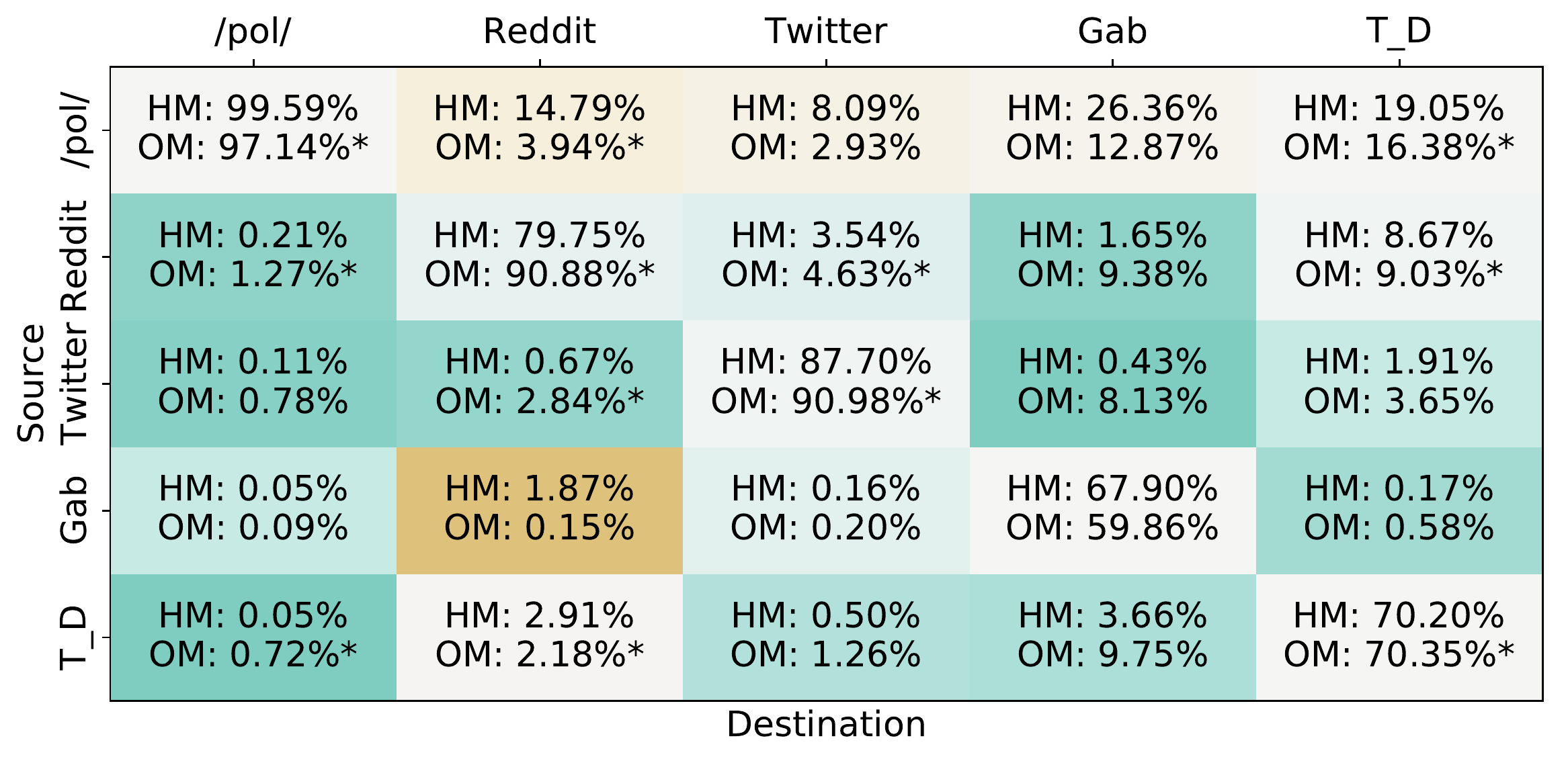}
\caption{Percent of the destination community's Happy Merchant (HM) and non-Happy-Merchant (OM) memes caused by the source community.  Colors indicate the percent difference between Happy Merchants and non-Happy-Merchants, while $*$ indicate statistical significance between the distributions with $p < 0.01$.}
\label{fig:hawkes_from}
\end{figure}

\begin{figure}[t!]
\centering
\includegraphics[width=\columnwidth]{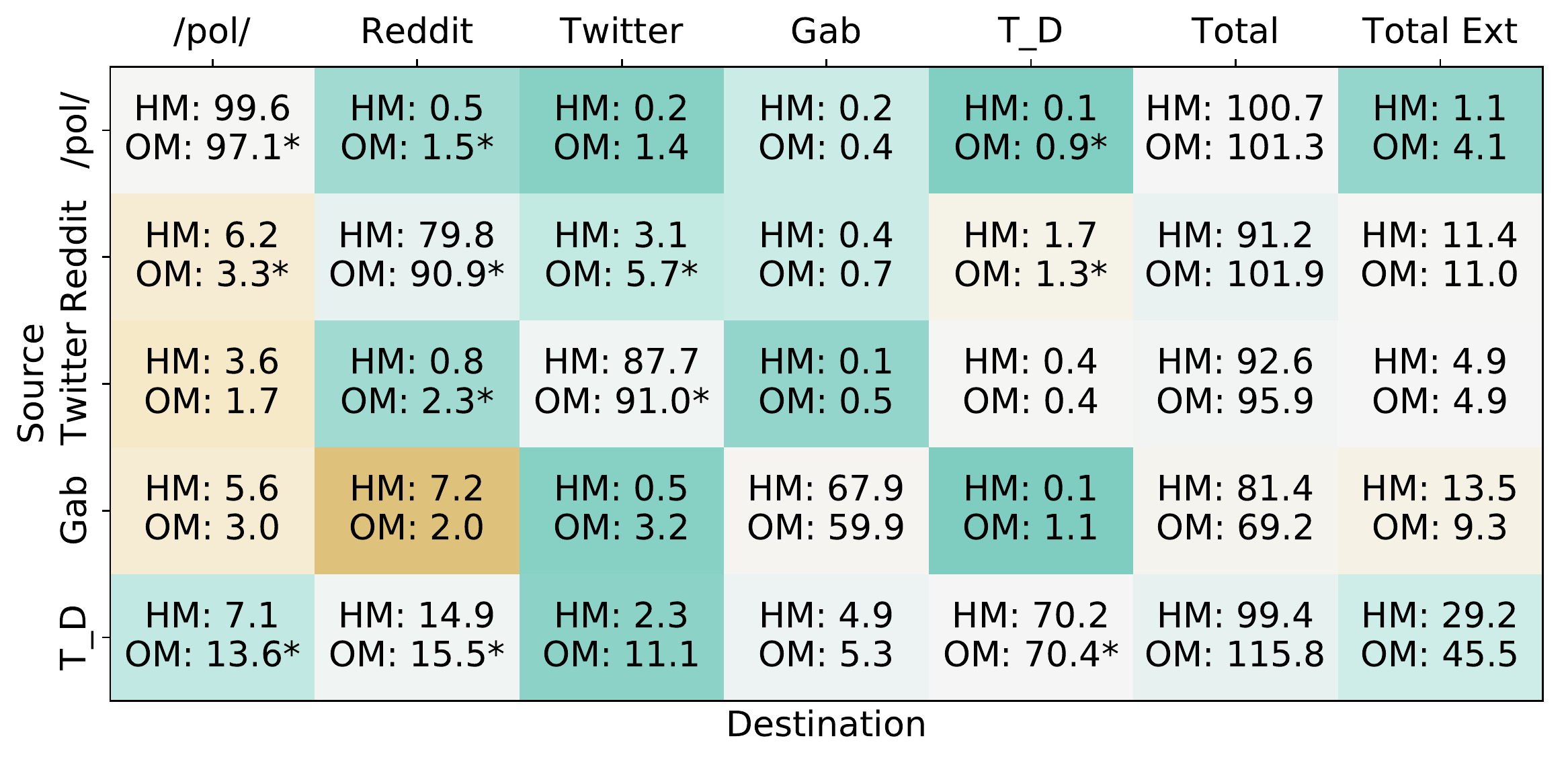}
\caption{Influence from source to destination community of Happy Merchant and non-Happy-Merchant memes, normalized by the number of events in the \textit{source} community, while $*$ indicate statistical significance between the distributions with $p < 0.01$.}
\label{fig:hawkes_norm}
\end{figure}

First, we report the percentage of events expected to be attributable from a source community to a destination community in Fig.~\ref{fig:hawkes_from}.
In other words, this shows the percentage of memes posted on one community which, in the context of our model, are expected to occur in direct response to posts in the source community. We can thus interpret this percentage in terms of the relative influence of meme postings one network on another.
We also report influence in terms of efficacy by normalizing the influence that each source community has, relative to the total number of memes they post (Fig.~\ref{fig:hawkes_norm}).
We compare the influence that Web communities exert on one another for the Happy Merchant memes (HM) and all other memes (OM) in the graph.
To assess the statistical significance of the results, we perform two-sample Kolmogorov-Smirnov tests that compare the distributions of influence from the Happy Merchant and other memes; an asterisk within a cell denotes that the distributions of influence between the source and destination platform have statistically significant differences ($p<0.01$).

Our results show that \dspol is the single most influential community for the spread of memes to all other Web communities.
Interestingly, the influence that \dspol exhibits in the spread of the Happy Merchant surpasses its influence in the spread of other memes.
However, although \dspol's overall influence is higher on these networks, its per-meme efficacy for the spread of antisemitic memes tended to be lower relative to non-antisemitic memes with the intriguing exception of \td.
Another interesting feature we observe about this trend is that memes on \dspol itself show little influence from other Web communities; both in terms of memes generally, and non-antisemitic memes in particular.
This suggests a unidirectional meme flow and influence from \dspol and furthermore, suggest that \dspol acts as a primary reservoir to incubate and transmit antisemitism to downstream Web communities.

\descr{Main Take-Aways.} To summarize, the main take-away points from our quantitative assessment are:
\begin{compactenum}
\item Racial and ethnic slurs are increasing in popularity on fringe Web communities. This trend is particularly notable for antisemitic language.
\item Our word2vec models in conjunction with graph visualization techniques, demonstrate an explosion in diversity of coded language for racial slurs used in \dspol and Gab.  Our methods demonstrate a means to dissect this language and decode racial discourse on fringe communities.
\item The use of ethnic and antisemitic terms on Web communities is substantially influenced by real-world events. For instance, our analysis shows a substantial increase in the use of ethnic slurs including the term ``jew'' around Donald Trump's Inauguration, while the same applies for the term ``white''  and the Charlottesville rally.
\item When it comes to the use of antisemitic memes, we find that \dspol consistently shares the Happy Merchant Meme, while for Gab we observe an increase in the use in 2017, especially after the Charlottesville rally. 
Finally, our influence estimation analysis reveals that \dspol is the most influential actor in the overall spread of the Happy Merchant to other communities, possibly due to the large volume of Happy merchant memes that are shared within the platform.  The\_Donald however, is the most efficient in pushing Happy Merchant memes to other Web communities. 
\end{compactenum}

\section{Discussion}\label{sec:discussion}

Antisemitism has been a historical harbinger of ethnic strife~\cite{adl_antisemitism_history,history_antisemitism_history}.
While organizations have been tackling antisemitism and its associated societal issues for decades, the rise and ubiquitous nature of the Web has raised new concerns.
Antisemitism and hate have grown and proliferated rapidly online, and have done so mostly unchecked.
This is due, in large part, to the scale and speed of the online world, and calls for new techniques to better understand and combat this worrying behavior.

In this paper, we take the first step towards establishing a large-scale quantitative understanding of antisemitism online.
We analyze over 100M posts from July, 2016 to January, 2018 from two of the largest fringe communities on the Web: 4chan's Politically Incorrect board (\dspol) and Gab.
We find evidence of increasing antisemitism and the use of racially charged language, in large part correlating with real-world political events like the 2016 US Presidential Election.
We then analyze the \emph{context} this language is used in via word2vec, and discover several distinct facets of antisemitic language, ranging from slurs to conspiracy theories grounded in biblical literature.
Finally, we examine the prevalence and propagation of the antisemitic ``Happy Merchant'' meme, finding that 4chan's \dspol and Reddit's \td are the most influential and efficient, respectively, in spreading this antisemitic meme across the Web.

Naturally our work has some limitations.
First, most of our results should be considered a \emph{lower bound} on the use of antisemitic language and imagery.
In particular, we note that our quantification of the use of the ``Happy Merchant'' meme is extremely conservative.
The meme processing pipeline we use is tuned in such a way that many Happy Merchant variants are clustered along with their ``parent'' meme.
Second, our quantification of the growth antisemitic language is focused on two particular keywords, although we also show how new rhetoric is discoverable.
Third, we focus primarily on two specific fringe communities.
As a new community, Gab in particular is still rapidly evolving, and so treating it as a stable community (e.g.,~Hawkes processes), may cause us to underestimate its influence.

Regardless, there are several important recommendations we can draw from our results.
First, organizations such as the ADL and SPLC should refocus their efforts towards open, data-driven methods.
Small-scale, qualitative understanding is still incredibly important, especially with regard to understanding offline behavior. 
However, resources \emph{must} be devoted to large-scale data analysis.
Second, we believe that--regardless of the participation of anti-hate organizations--scientists, and particularly computer scientists, must expend effort at understanding, measuring, and combating online antisemitism and online hate in general.
The Web has changed the world in ways that were unimaginable even ten years ago.
The world has shrunk, and the Information Age is in full effect.
Unfortunately, many of the innovations that make the world what it is today were created with little thought to their negative consequences.
For a long time, technology innovators have not considered potential negative impacts of the services they create, in some ways abdicating their responsibility to society.
The present work provides solid quantified evidence that the technology that has had incredibly positive results for society is being co-opted by actors that have harnessed it in worrying ways, using the same concepts of scale, speed, and network effects to greatly expand their influence and effects on the rest of the Web and the world at large.

\descr{Acknowledgments.} Savvas Zannettou received funding from the European Union's Horizon 2020 Research and Innovation program under the Marie Sk\l{}odowska-Curie ENCASE project (Grant Agreement No. 691025). We also gratefully acknowledge the support of the NVIDIA Corporation, for the donation of the two Titan Xp GPUs used for our experiments.

\small

\bibliographystyle{abbrv}

\end{document}